\documentclass{article}

\usepackage[position, final]{neurips_2026}
\usepackage{amsmath}
\usepackage{amsthm}
\usepackage{cancel}
\usepackage{amssymb}
\usepackage{amsfonts}
\usepackage{utfsym} % tick and x
\usepackage{footmisc}
\usepackage{hyperref} % disallowed by AAAI, duplicate with USENIX
\usepackage{url} % conflict with AAAI, duplicate with USENIX
\usepackage{graphicx}
\usepackage{multirow}
\usepackage{makecell}
\usepackage[flushleft]{threeparttable}
\usepackage{longtable}
\usepackage{colortbl}
\usepackage{booktabs}
\usepackage{subcaption}
\usepackage{xspace}
\usepackage{xcolor}
\usepackage{enumitem}
\usepackage[most]{tcolorbox}
\usepackage{fontawesome5}

\newcommand{\LLMBoxFontSize}{\scriptsize}

\tcbset{
  takeaway/.style={
    colback=gray!10,       % light gray background
    colframe=gray!60,      % darker gray border
    boxrule=0.8pt,         % border thickness
    arc=4mm,               % <-- round corners
    left=5mm, right=5mm,   % horizontal padding
    top=3mm, bottom=3mm,   % vertical padding
    fonttitle=\bfseries,
    coltitle=black
  }
}

\tcbset{
  llmcommon/.style={
    enhanced,
    breakable,
    sharp corners,
    boxrule=0.5pt,
    colback=white,
    top=4pt, bottom=4pt,
    left=6pt, right=6pt,
    before skip=8pt, after skip=8pt,
    fonttitle=\bfseries\sffamily\footnotesize,
    coltitle=black,
    fontupper=\LLMBoxFontSize,   % <- global font size applied to content
  }
}

\newtcolorbox{promptbox}[1][]{
  llmcommon,
  colback=blue!3,
  colframe=blue!70!white,
  fontupper=\LLMBoxFontSize\ttfamily,
  title={\faRobot\ LLM Prompt},
  #1
}

\newtcolorbox{llmfeedbackbox}[1][]{
  llmcommon,
  colback=green!3,
  colframe=green!60!black,
  fontupper=\LLMBoxFontSize\ttfamily,
  title={\faRobot\ LLM Feedback},
  #1
}

\newtcolorbox{humanreviewbox}[1][]{
  llmcommon,
  colback=yellow!4,
  colframe=yellow!50!black,
  title={\faUserEdit\ Reviewer Comment},
  #1
}

\newtcolorbox{instructionbox}[1][]{
  llmcommon,
  colback=orange!4,
  colframe=orange!70!black,
  title={\faInfoCircle\ Instructions},
  #1
}

\newtcolorbox{fullsummarybox}[1][]{
  llmcommon,
  colback=purple!5,
  colframe=purple!60!white,
  title={\faList\ Full Review Summary},
  #1
}

\newtcolorbox{metareviewbox}[1][]{
  llmcommon,
  colback=cyan!5,
  colframe=cyan!60!black,
  title={\faUsers\ AC Meta-review},
  #1
}
\title{Peer review should constrain evaluative authority}

\author{%
  Hanrui Wang\textsuperscript{1}\thanks{Corresponding author: \href{mailto:hanrui_wang@nii.ac.jp}{\texttt{hanrui\_wang@nii.ac.jp}}},~%
  Timo Spinde\textsuperscript{1,2},~%
  Isabella Habereder\textsuperscript{2,3},~%
  Chun-Shien Lu\textsuperscript{4},~%
  Isao Echizen\textsuperscript{1,2}
  \\
  \textsuperscript{1}National Institute of Informatics, Japan
  \\
  \textsuperscript{2}Media Bias Group, Japan
  \\
  \textsuperscript{3}Georg-August-University Göttingen, Germany
  \\
  \textsuperscript{4}Academia Sinica, Taiwan  
}

\begin{document}

\maketitle

\begin{abstract}
\textbf{Our position shifts the target of peer-review reform from reducing uncertainty through evaluator improvement or judgment aggregation to constraining evaluative authority under persistent uncertainty.}

Scientific peer review aims to assure research quality but also allocates scarce academic opportunities, such as publication in high-impact venues. However, the review system is under growing pressure as submissions increase while reviewer capacity, expertise, scrutiny, and incentives cannot scale accordingly. As peer review operates under persistent evaluative uncertainty, evaluator improvement and judgment aggregation remain structurally limited. We therefore propose authority-constrained peer review: before a consequential judgment is released as decision-relevant, it should withstand structured scrutiny through critique-authority decoupling, independent secondary evaluation, contestability and reversibility, and accountability for unsupported influence. Our aim is to prevent weakly supported judgments from entering final deliberation with the same authority as judgments that have withstood structured scrutiny, while creating incentives for reviewers to justify consequential claims more carefully. For large-scale computer-science conferences, this position offers an institutional response to uneven review quality and reviewer-assignment dependence; more broadly, it provides a governance principle for any domain where uncertain expert judgments allocate scarce academic opportunities.

\href{https://azrealwang.github.io/authority-constrained-peer-review/}{\textbf{Project Page}} summarizes this work and will be updated with future policy and technical developments related to peer review.
\end{abstract}

\section{Introduction}\label{introduction}

\begin{figure}[!ht]
    \centering
    \includegraphics[width=4.5in]{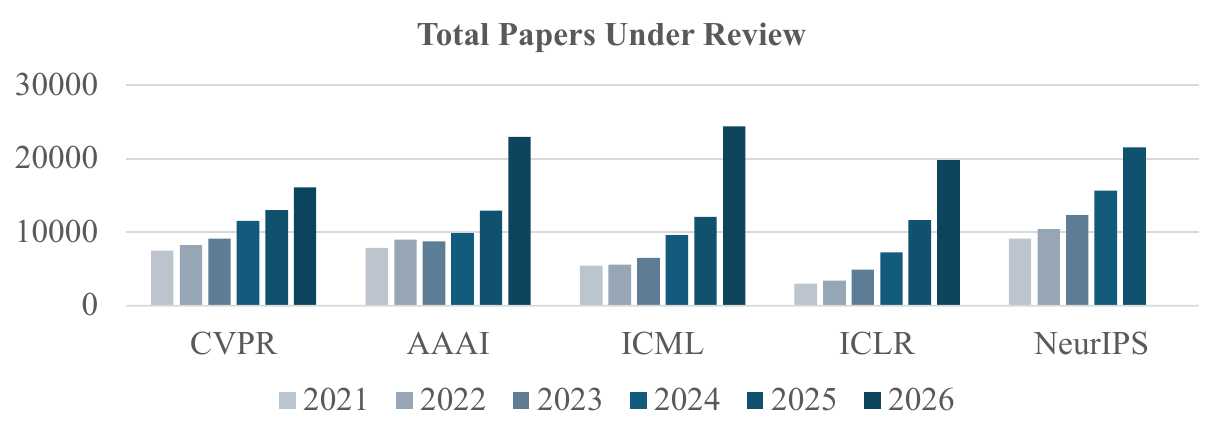}
    \caption{Total papers under review at major computer science venues across recent years, illustrating large growth in submission volume.}
    \label{fig_amount}
\end{figure}

Peer review is a central mechanism for ensuring the quality of scientific findings, but it also serves as a basis for allocating scarce opportunities for publication, funding, and professional recognition~\cite{siler2015measuring,aczel2025present}. However, in large-scale machine learning and AI conferences, the number of submissions under review has increased substantially in recent years (see Figure~\ref{fig_amount}). This growth not only increases reviewing workload, but also places greater pressure on review judgments: these judgments continue to shape consequential decisions~\cite{siler2015measuring,kravitz2010editorial} in a review system where reviewer expertise, review scrutiny, and reviewing incentives cannot scale at the same pace~\cite{gallo2016influence,huisman2017duration,tite2007peer,breuning2015reviewer,fox2017recruitment,seghier2025paying}.

One major response to rising review pressure is to improve evaluators themselves, based on the intuition that better evaluators lead to better reviews~\cite{aczel2025present,hesselberg2023reviewer,zhuang2025large}. Human-centered reforms seek better reviewer training, matching, incentives, transparency, and accountability~\cite{aczel2025present,leyton2024matching,seghier2025paying,kim2025position,yang2025position}. AI-centered reforms aim to improve review quality or accelerate peer review through model optimization, automated review generation, or hybrid human-AI workflows~\cite{chang2025treereview,lee2025role,zhuang2025large}. However, many criteria that peer review relies on, such as novelty and significance, lack reliable ground truth at the time of review~\cite{lee2015commensuration,hug2024referees,aczel2025present,oxley2025variability}: whether the review judgment is correct cannot be directly verified when decisions are made. Peer review, therefore, operates under persistent evaluative uncertainty. As a result, evaluator improvement lacks stable targets against which improvement can be guided~\cite{sizo2025defining,superchi2019tools}; even when a target can be specified, it remains difficult to validate whether overall review quality has been effectively improved~\cite{sizo2025defining,superchi2019tools,pier2018low}.

\textbf{Taking persistent evaluative uncertainty in peer review as a premise, we shift the institutional focus of reform: constraining the authority of each review judgment before it can influence the final decision.} We propose four principles:
\begin{itemize}[leftmargin=1cm]
    \item \emph{Critique-authority decoupling:} the evaluator who provides the substantive critique should not unilaterally convert that critique into a consequential decision signal, such as an acceptance or rejection recommendation.

    \item \emph{Independent secondary evaluation:} consequential review judgments should receive independent second-order checks for clarity, evidential support, and interpretive validity.

    \item \emph{Contestability and reversibility:} review judgments should remain open to challenge and, where necessary, revision or reversal.

    \item \emph{Accountability for unsupported influence:} reviews that significantly affect decisions but are later found to be poorly supported should remain visible in the record.
\end{itemize}

Our position also differs from the conventional many-signal peer review. Many-signal review relies on multiple reviewers, rebuttals, and decision layers to let judgments complement, correct, or compete with one another~\cite{pier2018low,hug2024referees}. However, each review judgment in a many-signal system can still enter final deliberation as one evaluator's unilateral judgment and become a consequential input to the final decision~\cite{hug2024referees,kravitz2010editorial}. In contrast, our proposed authority-constrained review requires consequential judgments to withstand structured scrutiny before they are released as decision-relevant signals. The aim of authority-constrained review is not to add more opinions for aggregation, but to prevent weakly supported judgments from entering final deliberation with the same authority as judgments that have withstood structured scrutiny. Constraining authority also creates incentives for reviewers to provide judgments that are more explicit, evidence-based, and defensible.

Taken together, our position does not rely on reducing evaluative uncertainty. Instead, it aims to prevent uncertain judgments from acquiring unchecked authority in final decisions and to encourage more careful review. More generally, it asks how peer-review institutions should govern the authority of expert judgments when those judgments allocate scarce opportunities under persistent uncertainty.

%%%%%%%%%%%%%%%%%%%%%%%%%%%%%%%%%%%%%%%%%%%%%%%%

\section{Core condition of peer review and its structural consequences}
\label{core}

% summarize criteria
Peer review relies on multiple evaluative criteria to reach final review decisions~\cite{lee2015commensuration,hug2024referees}. Table~\ref{tab_review_criteria} summarizes ten major criteria drawn from thirteen review guidelines.\footnote{\label{footnote_review_policies}\url{https://cvpr.thecvf.com/Conferences/2026/ReviewerGuidelines}\\\url{https://iccv.thecvf.com/Conferences/2025/ReviewerGuidelines}\\\url{https://eccv.ecva.net/Conferences/2026/ReviewerGuide}\\\url{https://icml.cc/Conferences/2026/ReviewerInstructions}\\\url{https://neurips.cc/Conferences/2025/ReviewerGuidelines}\\\url{https://iclr.cc/Conferences/2026/ReviewerGuide}\\\url{https://aaai.org/conference/aaai/aaai-26/instructions-for-aaai-26-reviewers/}\\\url{https://acmmm2025.org/reviewer-and-area-chair-guidelines/}\\\url{https://aclrollingreview.org/reviewerguidelines}\\\url{https://www.sigsac.org/ccs/CCS2026/call-for/call-for-papers.html}\\\url{https://signalprocessingsociety.org/publications-resources/guidelines-reviewers}\\\url{https://dl.acm.org/journal/tos/reviewer-guidelines}\\\url{https://www.science.org/content/page/science-information-reviewers}} 
% classify criteria
As presented in Section~\ref{introduction} and Table~\ref{tab_review_criteria}, many of these criteria lack reliable ground truth at the time decisions are made~\cite{lee2015commensuration,oxley2025variability}. Building on prior literature~\cite{macleod2021mdar,gallo2016influence,lee2015commensuration}, we classify these criteria into three \emph{ground truth statuses}, according to the extent to which reliable ground truth can be established at the time of review decisions. \emph{Relatively reliable:} criteria for which the relevant evidence is largely contained in the submission and can be directly assessed during review~\cite{macleod2021mdar}. \emph{Limited:} criteria for which some evidence is available during review, but assessment remains incomplete, conditional, or dependent on reviewer knowledge, scrutiny, or available resources~\cite{gallo2016influence}. \emph{Unavailable:} criteria for which no reliable ground truth can be established at decision time because the necessary evidence is not observable during review~\cite{lee2015commensuration}.
% restate core condition
We therefore affirm that missing criteria ground truth is the core condition under which peer review operates. This condition produces three structural consequences that are difficult to resolve, which together place peer review under persistent evaluative uncertainty.

\setlength{\cmidrulekern}{3em}
\aboverulesep=0.5pt
\belowrulesep=0.5pt

\begin{table*}[!ht]
\centering
\footnotesize
\caption{Major evaluative criteria in peer review and their ground truth (GT) status. Only a narrow subset admits relatively reliable decision-time ground truth. The classification reflects the availability of evidence, not perfect objectivity or mechanical decidability.}
\label{tab_review_criteria}
\setlength{\tabcolsep}{4pt}
\begin{threeparttable}
\begin{tabular}{p{2.8cm} p{2.1cm} p{8.2cm}}
\toprule
\textbf{Criterion} &
\textbf{GT status} &
\textbf{Note} \\
\midrule

Reporting completeness &
Relatively reliable &
Local omissions or reporting defects can be checked directly from the manuscript; mismatches may cause desk rejection~\cite{hopewell2025consort}.\\
\cmidrule{1-3}

Technical correctness / soundness &
Limited &
Some flaws can be identified during review, but deeper errors may remain undetected until later scrutiny~\cite{chauvin2015most}. \\
\cmidrule{1-3}

Methodological / empirical adequacy &
Limited &
Assessment depends on field standards, expertise, and depth of scrutiny~\cite{rodda2024peer}. \\
\cmidrule{1-3}

Literature review sufficiency &
Limited &
Coverage depends on reviewer knowledge, search depth, and judgments of relevance~\cite{mcgowan2016press}. \\
\cmidrule{1-3}

Reproducibility &
Limited &
Some evidence may be available at decision time, but promised code, data, or artifacts may later be incomplete or unavailable~\cite{nust2021codecheck}. \\
\cmidrule{1-3}

Ethics &
Limited &
Some issues are visible at review time~\cite{knight2024emerging}, whereas others may only be questioned or uncovered after publication~\cite{frank2023raising}. \\
\cmidrule{1-3}

Novelty / originality &
Unavailable &
Judgments depend on literature coverage, field framing, and what counts as a non-trivial advance~\cite{lee2015commensuration}. \\
\cmidrule{1-3}

Significance / importance / expected impact &
Unavailable &
Scientific importance or future impact cannot be reliably established at decision time~\cite{lee2015commensuration, glanzel2003better, ke2015defining}. \\
\cmidrule{1-3}

Venue fit / relevance &
Unavailable &
Assessment depends on venue identity and contestable expectations about audience and contribution.\tnote{a} \\
\cmidrule{1-3}

Clarity / presentation &
Unavailable &
Whether a paper is sufficiently clear for acceptance depends on reader background and expectations.\tnote{b} \\

\bottomrule
\end{tabular}
\end{threeparttable}
\begin{tablenotes}
  \scriptsize
    \item $^a$ \url{https://chi2025.acm.org/guide-to-reviewing-papers/}
    \item $^b$ \url{https://www.nature.com/nature/editorial-policies/peer-review/}
\end{tablenotes}
\end{table*}

\subsection{Standard inconsistency}
\label{standard_inconsistency}

% define reviewer-side subjectivity
Review guidelines are intended to ensure that the same evaluation criteria are applied to all submissions~\cite{hug2024referees}, but the standards by which reviewers interpret and apply these criteria remain heterogeneous~\cite{lee2015commensuration,hug2024referees}.
Building on prior literature \cite{leyton2024matching, luukkonen2012conservatism, weinberg2024reviewer, huisman2017duration, nust2021codecheck, hug2024referees, gallo2016influence}, we identify six \emph{reviewer-side sources of interpretive subjectivity} that shape how evaluative criteria are assessed (see Table~\ref{tab_subjectivity}). \emph{Topic familiarity:} familiarity with the paper's specific topic, problem setting, or literature~\cite{leyton2024matching}. \emph{Expertise:} relevant technical or methodological competence~\cite{gallo2016influence}. \emph{Scholarly pursuit:} the reviewer's orientation toward what kinds of research are worth advancing~\cite{luukkonen2012conservatism}. \emph{Preferences:} intellectual, methodological, or stylistic inclinations~\cite{weinberg2024reviewer}. \emph{Reviewer effort:} the time and attention invested in the review~\cite{huisman2017duration}.
\emph{Verification resources:} the practical means available to check relevant aspects of the submission~\cite{nust2021codecheck}.

% state consequence
Reviewers may interpret and apply the same criteria differently, making criteria ground truth difficult to determine collectively at the time of decision~\cite{pier2018low,hug2024referees}. Reviewer judgments may therefore diverge not only because submissions differ, but also because reviewers rely on different standards, evidence, and interpretive frames~\cite{pier2018low,hug2024referees,lee2015commensuration,gallo2016influence,oxley2025variability}. We therefore argue that peer review operates under standard inconsistency.

\begin{table*}[!ht]
\centering
\footnotesize
\caption{Illustrative sources of reviewer-side subjectivity in applying major evaluative criteria.
\textbf{TF}: topic familiarity; \textbf{EX}: expertise; \textbf{PU}: scholarly pursuit; \textbf{PR}: preferences; \textbf{EF}: reviewer effort; \textbf{RS}: verification resources.}
\label{tab_subjectivity}
\setlength{\tabcolsep}{11pt} % adjust inter-column spacing here
\newcommand{\critcol}{6cm} % adjust first-column width here
\begin{tabular}{p{\critcol}cccccc}
\toprule
\textbf{Criterion} & \textbf{TF} & \textbf{EX} & \textbf{PU} & \textbf{PR} & \textbf{EF} & \textbf{RS} \\
\midrule

Reporting completeness               &   &   &   &   & $\checkmark$ &   \\
\cmidrule{1-7}

Technical correctness / soundness   & $\checkmark$ & $\checkmark$ &   &   & $\checkmark$ & $\checkmark$ \\
\cmidrule{1-7}

Methodological / empirical adequacy & $\checkmark$ & $\checkmark$ & $\checkmark$ & $\checkmark$ &   &   \\
\cmidrule{1-7}

Literature review sufficiency       & $\checkmark$ & $\checkmark$ & $\checkmark$ & $\checkmark$ & $\checkmark$ &   \\
\cmidrule{1-7}

Reproducibility                     & $\checkmark$ & $\checkmark$ & $\checkmark$ &   & $\checkmark$ & $\checkmark$ \\
\cmidrule{1-7}

Ethics                              & $\checkmark$ & $\checkmark$ & $\checkmark$ & $\checkmark$ &   &   \\
\cmidrule{1-7}

Novelty / originality               & $\checkmark$ & $\checkmark$ & $\checkmark$ & $\checkmark$ &   &   \\
\cmidrule{1-7}

Significance / importance / expected impact
                                    & $\checkmark$ & $\checkmark$ & $\checkmark$ & $\checkmark$ &   &   \\
\cmidrule{1-7}

Venue fit / relevance               & $\checkmark$ & $\checkmark$ & $\checkmark$ & $\checkmark$ &   &   \\
\cmidrule{1-7}

Clarity / presentation              & $\checkmark$ &   & $\checkmark$ & $\checkmark$ &   &   \\

\bottomrule
\end{tabular}
\end{table*}

\subsection{Trust-dependent opacity}
\label{trust_dependent_opacity}

% define incentive conditions
Reviews may be shaped not only by how criteria are interpreted, but also by surrounding incentive conditions~\cite{bonn2021advancing,lim2025regional,trueblood2025misalignment,michelson2022association,resnik2018conflict,aczel2025present,bianchi2022can,tite2007peer,seghier2025paying}. We summarize three incentive conditions that are particularly relevant to peer review. \emph{Scarce rewards:} the fact that publication records are tied to prestige~\cite{bonn2021advancing}, career advancement~\cite{lim2025regional,trueblood2025misalignment}, visibility~\cite{bonn2021advancing}, and downstream access to resources~\cite{michelson2022association}. \emph{Conflict of interest (COI):} direct rivalry, shared interests, professional or network ties~\cite{resnik2018conflict}, as well as concerns about retaliation~\cite{aczel2025present,bianchi2022can}. \emph{Weak incentives for careful review:} the fact that reviewing is often time-consuming and only weakly rewarded~\cite{tite2007peer,seghier2025paying}.

% define incentive-coupled evaluations 
We refer to reviews shaped by such surrounding incentive conditions as \emph{incentive-coupled evaluations}. As discussed in the literature~\cite{buenstorf2025analyzing,resnik2018conflict,righi2017miracle,kocak2025ensuring,bianchi2022can}, incentive-coupled evaluations may take at least three forms. \emph{Strategic distortion:} deliberate inflation or deflation of evaluative judgments, for example, through harsher or more favorable assessments of novelty or significance; such distortion may be motivated by scarce rewards or COI~\cite{buenstorf2025analyzing,resnik2018conflict}.  \emph{Low-effort or AI-substituted review:} evaluation based on limited effort or AI substitution, leading to superficial comments, heuristic judgment, or weakly scrutinized review text; this may arise when reviewers prioritize research production over careful reviewing under weak reviewing incentives~\cite{righi2017miracle,kocak2025ensuring}. \emph{Cautious non-commitment:} hedged or persistently borderline judgment that avoids strong commitment, often reflecting limited evaluative investment or reluctance to impose a clear acceptance or rejection; such judgment may also be reinforced by COI, particularly concerns about retaliation~\cite{bianchi2022can}.

% Bringing it together and stating the consequence
In Table~\ref{tab_incentive_coupling}, we illustrate couplings between incentive conditions and forms of incentive-coupled evaluation. Because many review criteria have limited or unavailable ground truth, incentive-coupled evaluations are difficult to detect or separate from ordinary criteria-based disagreement~\cite{resnik2018conflict,lee2015commensuration}. Peer review, therefore, relies partly on trust in reviewers' professional and ethical conduct.

\begin{table*}[!ht]
\centering
\footnotesize
\caption{Illustrative couplings between incentive conditions and incentive-coupled evaluations.}
\label{tab_incentive_coupling}
\setlength{\tabcolsep}{8pt}
\begin{tabular}{p{5cm}ccc}
\toprule
\textbf{Incentive-coupled evaluation} & \textbf{Scarce rewards} & \textbf{COI} & \textbf{Weak incentives for careful review} \\
\midrule

Strategic distortion & $\checkmark$ & $\checkmark$ &   \\
\cmidrule{1-4}

Low-effort / AI-substituted review & $\checkmark$ &   & $\checkmark$ \\
\cmidrule{1-4}

Cautious non-commitment &   & $\checkmark$ & \\

\bottomrule
\end{tabular}
\end{table*}

\subsection{Criteria-to-decision indeterminacy}
\label{criteria_to_decision_indeterminacy}

\begin{figure}[!ht]
    \centering
    \includegraphics[width=5in]{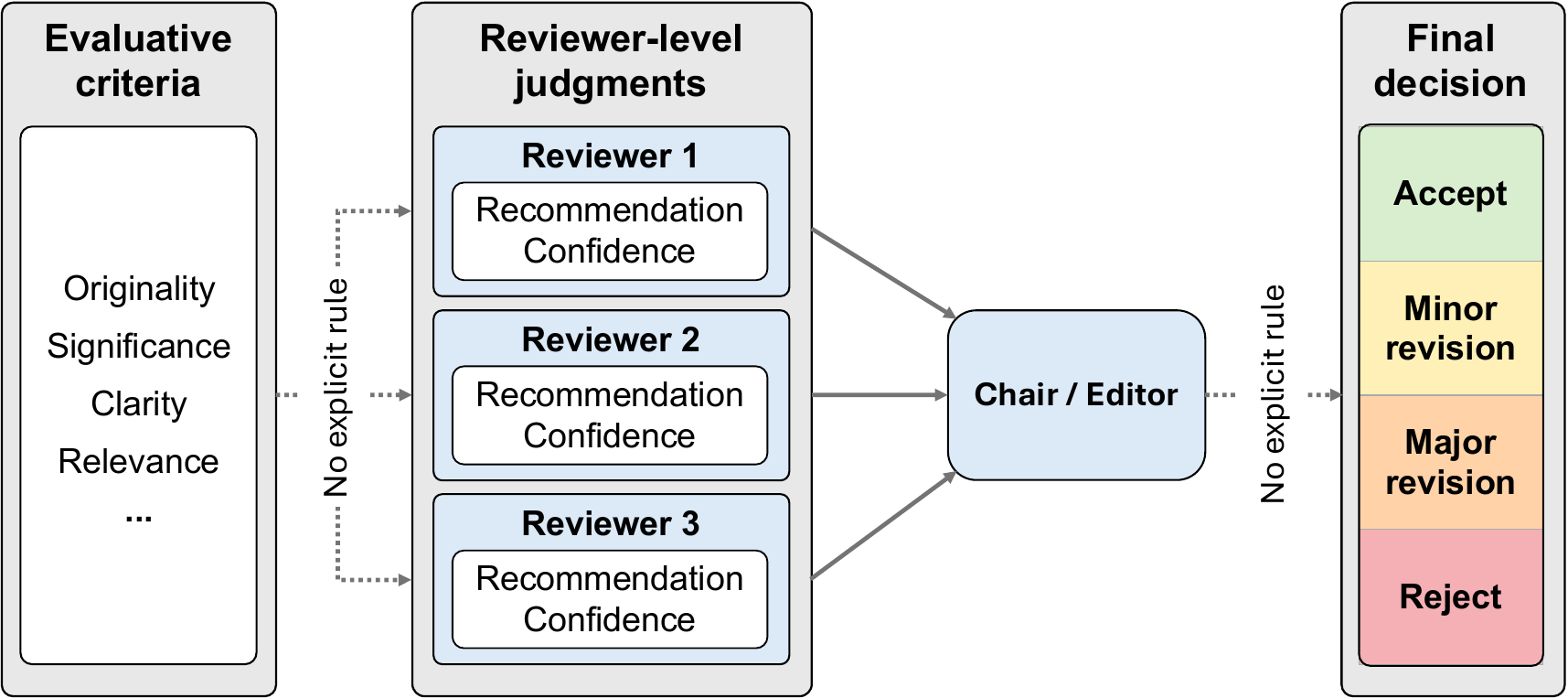}
    \caption{Peer-review criteria indirectly inform reviewer-level recommendation and confidence judgments, which, together with evaluative comments, are then further considered by chairs or editors in reaching a final decision. No explicit rule links criteria to reviewer judgments or reviewer judgments to final decisions.}
    \label{fig_criteria_to_decision}
\end{figure}

% describe peer review pipeline
As illustrated in Figure~\ref{fig_criteria_to_decision}, peer review involves a two-stage process: evaluative criteria (Table~\ref{tab_review_criteria}) are first translated into reviewer-level judgments (e.g., recommendations and confidence scores), which are then used by editors or chairs to reach final decisions.\footref{footnote_review_policies}

However, neither stage follows explicit or stable decision rules, i.e., there is no standardized way to translate criteria into reviewer judgments, nor to translate those judgments into final decisions~\cite{hug2024referees,kravitz2010editorial}. As a result, evaluations that use the same criteria may still yield different outcomes. We refer to this structural lack of a well-defined mapping as \emph{criteria-to-decision indeterminacy}.

%%%%%%%%%%%%%%%%%%%%%%%%%%%%%%%%%%%%%%%%%%
\section{Alternative views}
\label{alternatives}

As stated in Section~\ref{introduction}, we examine three prominent approaches to improving peer review: many-signal review, human-centered improvement, and AI-assisted review. The following sections analyze what each approach can improve and why reducing uncertainty through judgment aggregation or evaluator improvement remains insufficient under missing criteria ground truth.

\subsection{Many-signal review systems}
\label{reform_multiplicity}

% Aim/Goal of many-signal review 
Conventional peer review often uses multiple reviewers, author rebuttals, and layered decision-making by chairs and editors; we refer to this as a many-signal review system.\footnote{\label{footnote_review_policy}\url{https://www.acm.org/publications/policies/roles-and-responsibilities}} The intuition of many-signal review is that distributing judgment across reviewers and decision layers can reduce the impact of a single reviewer's idiosyncratic judgment on final decisions~\cite{pier2018low,hug2024referees}. 

% Weakness 1: any single review is still possible to be problematic, and can anchor later decision
However, each reviewer in a many-signal system still evaluates under missing criteria ground truth, applies heterogeneous standards, may be influenced by surrounding incentive conditions, and contributes to a decision process without a stable criteria-to-decision mapping. Adding more reviewers, therefore, does not prevent a single review from exerting consequential downstream influence~\cite{kravitz2010editorial,d2017can}. This influence may become significant when negative or cautionary interpretations carry disproportionate weight in subsequent decision-making~\cite{rozin2001negativity}, for example, because early judgments may anchor later evaluation~\cite{liu2024testing,tversky1974judgment}, or because competitive selection settings may make adjudicators more cautious toward uncertain work~\cite{luukkonen2012conservatism,gallo2018risk}.

A further limitation concerns the timing of correction. Many-signal systems allow a judgment to be challenged through other reviews, author rebuttal, reviewer discussion, or chair/editor moderation\footref{footnote_review_policy}, but such correction may occur after the judgment has already begun to shape the decision process~\cite{kravitz2010editorial,liu2024testing}. The key distinction is therefore not whether a judgment can eventually be corrected, but whether it is independently scrutinized before or only after it begins to exert consequential influence.

% Weakness 2: The number of reviewers in the current system is insufficient
Moreover, substantial disagreement among reviewers remains common under current many-signal review systems~\cite{hesselberg2025registered,aczel2025present}, suggesting that the number of reviewers typically assigned to a submission is often insufficient to stabilize evaluative variance. Existing studies do not clearly establish how many reviewers would be needed to reduce such variance substantially, nor do they show that greater agreement among a limited number of reviewers would guarantee correctness~\cite{pier2018low,erosheva2021zero,fogelholm2012panel}. Furthermore, the problem of reviewer disagreement is further compounded when review quality is degraded by incentive-coupled evaluations, for example, through low-effort review or cautious non-commitment.

\subsection{Human-centered evaluator improvement}
\label{reform_human}

% Aim/Goal of human-centered reform + back to our core condition
Human-centered reforms typically aim to improve peer review through clearer reviewing standards and training~\cite{hesselberg2023reviewer,yang2025position}, improved reviewer matching~\cite{shah2022challenges,jakobsen2022factors}, stronger transparency and accountability mechanisms~\cite{yang2025position,kim2025position}, and stronger incentives or expansion of the reviewer pool~\cite{seghier2025paying,kim2025position}. Their intuition is that peer review can be improved by better human reviewers and more robust review procedures. However, there is no universally accepted reference standard for a ``good'' review~\cite{sizo2025defining,superchi2019tools}, leaving evaluator improvement without a reliable target.

% Weakness 1: Can not eliminate reviewer side subjectivity
Improved training, matching, and procedural guidance may reduce some forms of inconsistency~\cite{hesselberg2023reviewer,jakobsen2022factors,leyton2024matching}, but they cannot remove differences in topic familiarity, expertise, scholarly pursuit, and evaluative preferences~\cite{leyton2024matching,gallo2016influence,luukkonen2012conservatism,weinberg2024reviewer}. Judgments about novelty, significance, and fit therefore remain contestable rather than reducible to uniform application of a shared standard~\cite{lee2015commensuration,li2025developing,hug2024referees}.

% Weakness 2: No meaningful compliance
Furthermore, improved review standards and procedures do not ensure meaningful compliance. Greater transparency, accountability requirements, and procedural expectations may encourage improved review conduct~\cite{yang2025position,kim2025position,hesselberg2023reviewer}, but they do not ensure that reviewers will follow the guidelines consistently in practice~\cite{bianchi2022can,resnik2018conflict,aczel2025present}. Reviewer overload~\cite{smith2006peer,aczel2025present}, difficulty securing willing reviewers~\cite{tite2007peer,breuning2015reviewer,fox2017recruitment}, and weak incentives~\cite{seghier2025paying,aczel2025present} continue to place reviewers under conflicting constraints of time, effort, and care.

% Weakness 3: Not scalable
Moreover, there is evidence of a sustained increase in the number of submissions (see Figure~\ref{fig_amount}), while it is unlikely that a human-centered reform would lead to a corresponding increase in reviewer capacity. Stronger incentives and author-obligated reviewing may increase reviewer supply, but they cannot generate scarce expertise on demand~\cite{gallo2016influence,bergstrom2026screening} or eliminate reviewer-submission mismatch~\cite{leyton2024matching,jakobsen2022factors,rogers2023program}. Expertise, carefulness, and institutional professionalism are difficult to expand at the pace of rising submission volume and AI-assisted research production~\cite{aczel2025present,zhang2025exploring,lu2026towards}.

\subsection{AI-assisted review}
\label{reform_ai}

% Aim/Goal of the AI-assisted review
AI-centered reforms typically aim to improve peer review through automated review generation~\cite{chang2025treereview,bougie2025generative,zhu2025deepreview}, AI-assisted review support~\cite{lee2025role,choi2026position,zhuang2025large}, and hybrid human-AI evaluation workflows~\cite{zhuang2025large}. Their central intuition is that peer review can be improved through scalable evaluative support that expands review capacity, increases consistency, and enhances the quality of critique~\cite{zhuang2025large}.

% Weakness 1: training and optimizing without clean data and target
However, review corpora~\cite{kuznetsov2022revise,lin2023moprd,kang2018dataset,dycke2023nlpeer} encode historically produced judgments shaped by standard inconsistency and trust-dependent opacity, rather than verified targets of correctness, significance, or future value~\cite{lee2015commensuration,glanzel2003better,ke2015defining}. Synthetic data may broaden the training signal~\cite{sadallah2025good}, but optimization and validation still rely on heterogeneous human judgments, evaluative standards, and heuristic metrics rather than verified criteria ground truth~\cite{pier2018low,lee2015commensuration,kirtani2025revieweval,glanzel2003better,ke2015defining}. AI-assisted review, therefore, inherits the core condition of missing criteria ground truth at the level of data, optimization target, and validation.

% Exerbation
Moreover, training and optimization of AI review systems become more challenging as the review stream itself is affected by variability and distortion. The missing criteria ground truth already limits the possibility of clean supervision. Reviewer-side subjectivity and incentive conditions further shape the evaluative signals that enter review corpora, while accelerated scientific output, limited reviewer capacity, and scarce rewards may accelerate the accumulation of inconsistency, bias, and signal degradation in these data~\cite{kocak2025ensuring,lo2025personal,liang2024monitoring}. AI reviews thus face a recursive problem: the data from which these systems learn are themselves produced under the very structural constraints that these systems are expected to overcome.

% Weakness 2: AI-peer-review-specific limitations
Existing studies document critical limitations of AI review, including factual unreliability~\cite{zhou2024llm,huang2025survey}, shallow assessment of scientific validity and technical depth~\cite{lee2025role,choi2026position,zhou2024llm}, and security, privacy, bias, and governance concerns~\cite{hosseini2023fighting,lin2025breaking,kousha2024artificial,zhuang2025large,ghaffar2026ethical}. These limitations are relevant but not central to our argument, as many may be partially mitigated by stronger models and more mature institutional safeguards~\cite{kousha2024artificial,lee2025role,zhuang2025large}. Our central concern is instead structural: model improvement may reduce generic LLM weaknesses, but it cannot, by itself, establish reliable performance on the specialized task of peer-review evaluation without a valid reference standard for review quality.

%%%%%%%%%%%%%%%%%%%%%%%%%%%%%%%%%%%%%%%%%%
\section{Our position: Authority-constrained peer review}
\label{design-impl}

We argued that existing approaches share the same problematic logic: they seek better decisions through more signals, better human reviewers, or AI evaluators, without reliable criteria ground truth. Under this logic, uncertain judgments remain difficult to distinguish from well-supported ones, yet uncertain judgments can still acquire unchecked evaluative authority in final deliberation. In contrast, we take such persistent evaluative uncertainty due to missing ground truth, including standard inconsistency, trust-dependent opacity, and criteria-to-decision indeterminacy, as a premise and shift the reform target from improving evaluators to governing evaluative authority. We use evaluative authority to refer to the capacity of a review judgment, score, objection, or recommendation to shape final deliberation or appear in the decision rationale.

\subsection{Design principles}

Authority-constrained review can be implemented through four principles. First, critique and authority should be decoupled: the evaluator who writes the substantive critique should not automatically convert it into a consequential recommendation or score. Second, consequential judgments should receive independent second-order evaluation of their clarity, evidential support, and interpretive validity before they enter the decision rationale. This is a targeted check of the existing review judgment, not an independent full review of the submission. Third, review judgments should remain contestable and reversible, so that challenges can lead to clarification, narrowing, downgrading, score revision, or removal from the decision rationale. Fourth, unsupported influence should remain accountable: when a judgment significantly shaped a decision but is later found to be poorly supported, the system should record both the judgment and the correction of its authority. These principles make review influence conditional rather than automatic, while encouraging reviewers to make consequential claims more carefully.

% emphasize differ from many signal
Authority-constrained review differs from many-signal review systems not only in how judgments are checked, but also in when they acquire consequential authority. Many-signal review adds signals for aggregation and may correct judgments through later rebuttal, discussion, or moderation, whereas authority-constrained review requires independent scrutiny before a judgment enters final deliberation with full decision relevance. The aim is to expose weak evidential bases, unsupported claims, and unresolved interpretive problems before they exert consequential influence, making authority conditional on prior scrutiny rather than later correction.

\subsection{Comparison with alternative views}

To clarify the distinctiveness of our position, Table~\ref{tab_strategy_compare} summarizes how authority-constrained peer review governs evaluative authority under the core condition and structural consequences developed in Section~\ref{core}, in contrast to the three approaches discussed in Section~\ref{alternatives}. We elaborate on each comparison below.

\begin{table*}[!ht]
\centering
\footnotesize
\caption{How different approaches address the core condition and its structural consequences.}
\label{tab_strategy_compare}
\setlength{\tabcolsep}{3pt}
\begin{tabular}{p{2.5cm} p{2.1cm} p{2.1cm} p{2.1cm} p{4cm}}
\toprule
\textbf{Core Condition / Consequence} & \textbf{Many-signal} & \textbf{Human-centered} & \textbf{AI-assisted} & \textbf{Authority-constrained} \\
\midrule
Missing criteria ground truth
& Add more signals
& Improve evaluators
& Optimize models on review data
& Assume persistence; limit decisional influence \\
\cmidrule{1-5}
Standard inconsistency
& Average across judgments
& Standardize human interpretation
& Standardize model scoring
& Cross-check interpretation before influence \\
\cmidrule{1-5}
Trust-dependent opacity
& Reduce single-review influence
& Improve detection and compliance
& Improve detection and filtering
& Constrain authority to make incentive-coupled influence harder \\
\cmidrule{1-5}
Criteria-to-decision indeterminacy
& Add evaluative inputs
& Clarify procedures
& Learn existing mappings
& Limit unilateral steering under indeterminacy \\
\bottomrule
\end{tabular}
\end{table*}

\textbf{Core condition.} Many-signal, human-centered, and AI-centered approaches are all oriented toward bringing peer review closer to the criteria ground truth: by aggregating more signals, by training better human reviewers, or by optimizing models on review data and preferences. In contrast, our position begins with the assumption that weakly supported, seriously flawed, or otherwise contestable reviews will persist in the absence of criteria ground truth and cannot be eliminated with confidence. We therefore seek to constrain the downstream influence of consequential judgments through independent scrutiny rather than to rely on directly improving their quality. This principle applies to all consequential judgments, although its practical implementation may need to allocate scrutiny unevenly. In an initial or resource-constrained deployment, judgments with greater potential influence on the final decision should therefore be checked more carefully, for example, clear accept or reject judgments that, if unsupported, could substantially shape the eventual outcome.

\textbf{Standard inconsistency.} Standard inconsistency arises primarily from reviewer-side human subjectivity, which cannot be reliably eliminated. Many-signal review responds by adding evaluators under the assumption that a larger number of judgments will reduce variance. Human-centered reform provides clearer, more detailed criteria to ensure more uniform interpretation. However, reviewer pools are usually too small for multiplicity alone to neutralize heterogeneous interpretation. Detailed standards also remain subject to divergent human understandings. Compared with many-signal and human-centered approaches, AI-assisted evaluation can respond more directly to standard inconsistency. A single model may apply a more consistent scoring standard across submissions. But consistent scoring does not imply that AI-assisted review is an optimal solution, given the additional limitations discussed in Section~\ref{reform_ai}. 
Instead of relying on multiple separate reviews, we propose a different form of evaluation diversity, in which evaluators cross-check one another's interpretations within the same review process. Our position is particularly clear in a human-critique, AI-score arrangement, in which the human evaluator's decision-making authority is limited, and the AI evaluator helps ensure greater consistency in scoring.

\textbf{Trust-dependent opacity.} Trust-dependent opacity arises because incentive-coupled evaluations cannot be cleanly ruled out under missing criteria ground truth. In a many-signal review system, it is assumed that including additional reviewers mitigates the influence of individual judgments driven by incentives. Human-centered and AI-centered reforms, by contrast, depend more on correcting and mitigating problematic evaluations through improved norms, supervision, or model design. We argue that, given missing criteria ground truth, detection of incentive-coupled evaluations rarely produces evidence strong enough to resolve the issue institutionally. Instead, our position of constrained authority does not aim to eliminate incentive-coupled evaluations, but makes them harder to shape outcomes without resistance.

\textbf{Criteria-to-decision indeterminacy.} Criteria-to-decision indeterminacy is the consequence least directly resolved by existing approaches, including our own. Many-signal review adds evaluative inputs but does not establish a stable guideline for aggregating criteria or reviewer-level judgments into final decisions. Human-centered reform can clarify criteria and procedures, but does not yield a determinate mapping from substantive evaluation to recommendations or from recommendations to final decisions. While AI-centered reform can learn or reproduce historical classifications from previous reviews and decisions, this merely mimics existing patterns rather than eliminating the underlying uncertainty. Our position does not eliminate the criteria-to-decision indeterminacy either. Instead, we limit the extent to which a single review can steer final decisions. In doing so, we limit the extent to which the criteria-to-decision indeterminacy from a single review can affect outcomes.

%%%%%%%%%%%%%%%%%%%%%%%%%%%%%%%%%%%%%%%%%%%%%
\section{One possible instantiation}
\label{inst-arrangement}

Authority-constrained review can be instantiated in different institutional forms. We describe one concrete illustration based on dual-track human--AI mutual supervision. In \emph{Track A}, an AI system produces a substantive critique without assigning a consequential score, and human supervisors independently assess the critique by recording agreement, disagreement, and important omissions before forming their own judgments. In \emph{Track B}, human reviewers produce substantive critiques, while an AI system independently assigns the rating most consistent with each critique and scrutinizes its evidential support, interpretive validity, and decision proportionality. The two tracks do not observe each other's intermediate outputs, and their outputs are provided to the chair or editor for final adjudication rather than automatically aggregated into a decision. This separation operationalizes the principle that the evaluator producing a substantive critique should not unilaterally determine its consequential authority.

A central component of this instantiation is \emph{structured scrutiny}. Rather than asking a second evaluator to produce another full review, scrutiny targets the basis on which an existing judgment could acquire decision influence. As summarized in Table~\ref{tab:scrutiny}, we distinguish three dimensions that assess whether a judgment is supported, correctly interpreted, and proportionate to its intended decision consequence, without requiring a single ground-truth evaluation.

\begin{table}[!ht]
\centering
\caption{Three dimensions of structured scrutiny for consequential review judgments.}
\label{tab:scrutiny}
\footnotesize
\setlength{\tabcolsep}{3pt}
\begin{tabular}{p{3cm} p{10.5cm}}
\toprule
\textbf{Dimension} & \textbf{Core question} \\
\midrule
Evidential support & Is the consequential claim factually supported by the submission or other relevant evidence? \\
Interpretive validity & Does the judgment correctly interpret the submission and the issue being evaluated? \\
Decision proportionality & Does the identified issue justify the amount of decision influence assigned to it? \\
\bottomrule
\end{tabular}
\end{table}

These three dimensions also provide a basis for interpreting agreement and disagreement. Agreement provides additional support for a judgment, whereas disagreement indicates that it requires closer examination. In the case of disagreement, the chair or editor can first compare the competing evaluations in terms of the three scrutiny dimensions. If needed, independent reviews and the relevant manuscript content can provide further evidence for adjudication. If the disagreement remains unresolved, the judgment may receive reduced authority, require clarification, or be prevented from determining the outcome by itself. Importantly, agreement does not establish that a review is correct, just as disagreement does not establish that it is wrong. Our position aims to prevent unresolved consequential judgments from exercising unchecked authority, rather than to claim that scrutiny signals reveal a ``more correct'' evaluation.

One concern is that a dual-track design may increase human reviewer workload. The workload implications, however, depend on the institutional realization. An all-human implementation introduces additional human scrutiny, although checking an existing judgment can be lighter than conducting another full review. In the AI--human realization studied here (Track A), some full human reviewing can instead be replaced by supervision of an existing AI critique, potentially resulting in a lighter reviewer-side workload. Chair and editor workload remains uncertain: more structured signals must be considered, but unsupported or conflicting judgments may also become easier to identify. We analyze the workload implications of different realizations in Appendix~\ref{workload_math}.

To examine the operational feasibility of the dual-track human--AI mutual supervision design, we instantiated both tracks on one real borderline submission using four original peer reviews, four recruited human supervisors, one experienced area chair, and ChatGPT-5.1. The study provides preliminary evidence for several aspects of the design. In Track A, all four human supervisors reported that supervising the AI-generated critique was easier than conventional reviewing, including two who voluntarily consulted the original manuscript. In Track B, AI scrutiny produced explicit counter-signals by identifying score--reasoning inconsistencies and potential review-quality concerns, while human and AI judgments also diverged in some cases. These counter-signals remained visible for area-chair adjudication rather than being resolved automatically. The area chair reported neither a clear increase nor decrease in adjudication difficulty, consistent with the uncertain chair/editor workload discussed above. Detailed settings, results, participant feedback, and limitations are provided in Appendix~\ref{toy}.

This feasibility study establishes operational plausibility rather than venue-scale effectiveness. It shows that the proposed role separation, structured scrutiny, reciprocal checking, and chair-level adjudication can be instantiated in a complete review process, but evaluating broader effects requires institutionally representative studies across reviewer populations, research fields, incentives, policies, and decision contexts. Building on these open questions, we outline a roadmap for developing and validating authority-constrained review, including improved scrutiny design, suitable peer-review datasets, more reliable counter-signal generation and adjudication, less-biased AI support, safeguards against technical misconduct, and institutionally representative pilot studies. These objectives are detailed in Appendix~\ref{objectives}.

%%%%%%%%%%%%%%%%%%%%%%%%%%%%%%%%%%%%%%%%%%%%%
\section{Scope}

First, this paper does not argue that scientific evaluation should cease because of uncertainty, nor that current peer review is broken. Our argument is institutional rather than accusatory: under present conditions, peer-review reform cannot safely rely solely on selecting trusted evaluators.

Second, this paper does not argue that AI is ineffective or that improving evaluators is futile. Better reviewer training, incentives, and AI systems may all improve the signals available to the system and, in some settings, substantially improve decision quality. However, such gains should be understood as improvements to evaluative inputs rather than as solutions to evaluative authority. Better signals do not, by themselves, justify unconstrained decision influence.

Third, this paper does not argue that expertise is irrelevant or that all signal sources should be treated identically. Expertise matters, and final responsibility cannot be eliminated. Our claim is narrower and more practical: under missing criteria ground truth, expertise alone should not exempt consequential judgments from structured checking, explicit challenge, and possible reversal.

Fourth, reducing reviewer workload is not the primary claim of authority-constrained review, although it is an important consideration in its practical realization. The central aim is to govern how consequential evaluative influence is exercised under missing criteria ground truth. Workload implications depend on the institutional design, and our human--AI instantiation illustrates how this governance objective can be pursued while also reducing reviewer-side workload.

\section{Conclusion}

Peer review operates under missing criteria ground truth: many criteria that shape consequential decisions cannot be verified against stable standards at the time of review. This condition limits approaches that seek better decisions primarily by adding signals, improving evaluators, or scaling evaluative support, because these approaches do not by themselves prevent uncertain or weakly supported judgments from acquiring unchecked authority in final deliberation. Our position, therefore, departs from efforts to reduce uncertainty and instead governs the conditions under which review judgments become decision-relevant. Its aim is to ensure that consequential judgments are exposed to scrutiny, challenge, and possible correction before they carry full authority in final decisions. For large-scale computer-science conferences, this position addresses the downstream effects of variable review quality and reviewer-assignment dependence.

More broadly, our argument in this position paper becomes more urgent as AI agents accelerate research production, manuscript preparation, and review assistance. The pressures now visible in computer-science conferences may therefore foreshadow challenges that other fields will face. Computer science is not only a domain under review pressure, but also an early testbed for institutional reform. Venues such as NeurIPS, where scale, AI-assisted research, and peer-review governance already intersect, are well positioned to pilot authority-constraining mechanisms. The practical task is not to wait for perfect evaluators, complete ground truth, or fully settled policies, but to test and refine institutional designs that make consequential review authority more accountable under persistent uncertainty.

%%%%%%%%%%%%%%%%%%%%%%%%%%%%%%%%%%%

%%%%%%%%%%%%%%%%%%%%%%%%%%%%%%%%%%%

%%%%%%%%%%%%%%%%%%%%%%%%%%%%%%%%%%%%%%%%
\begin{ack}
This work was partially supported by JSPS KAKENHI Grant JP24H00732, by JST CREST Grants JPMJCR20D3 and JPMJCR2562 including AIP challenge program, and by JST K Program Grant JPMJKP24C2 Japan. It was also funded by the German Federal Ministry of Education and Research (BMBF) through the DAAD (German Academic Exchange Service) and a fellowship by the Hanns-Seidel-Foundation. It was also supported by the Taiwan National Science and Technology Council (NSTC) with Grant 114-2634-F-001-001-MBK, and by the Deutsche Forschungsgemeinschaft (DFG, German Research Foundation) under project number 565115197 (BARI).  
\end{ack}
{
\small
% \nocite{*}
\bibliographystyle{unsrt}
\bibliography{references}
}
\newpage
\appendix

\section{Human Workload Analysis and Comparison Across Peer-Review Designs}
\label{workload_math}

The workload implications of authority-constrained review depend on its institutional realization. Compared with conventional independent human review, an all-human realization of mutual supervision increases human workload, whereas the dual-track human--AI realization described in Section~\ref{inst-arrangement} can reduce reviewer-side human workload. Table~\ref{tab:workload_roles} summarizes the workload implications across designs and roles.

\begin{table}[!ht]
\centering
\caption{Expected workload effects in a dual-track example.}
\label{tab:workload_roles}
\footnotesize
\setlength{\tabcolsep}{28pt}
\begin{tabular}{p{3cm} p{7cm}}
\toprule
\textbf{Role} & \textbf{Workload} \\
\midrule
Human reviewers
& Fewer full reviews \\

Human supervisors
& Lighter than writing a review from scratch \\

AI
& Additional computation with little human reviewing labor \\

Chair / editor
& Uncertain: more signals, but with greater structure \\
\bottomrule
\end{tabular}
\end{table}

\paragraph{Conventional independent human review.}
Consider a conventional design in which four human reviewers independently evaluate each submission. Let $W_{\mathrm{full}}$ denote the workload required for one full human review. The reviewer-side human workload is
\[
    W_{\mathrm{conv}} = 4W_{\mathrm{full}}.
\]
This provides the baseline against which we compare realizations of mutual supervision.

\paragraph{All-human mutual supervision.}
Authority-constrained review does not depend on AI adoption. One possible realization retains four full human reviewers and assigns an independent human supervisor to scrutinize each review judgment, yielding four final review judgments. Let $W_{\mathrm{sup}}$ denote the workload required to perform this second-order scrutiny. Reviewer-side human workload then becomes
\[
    W_{\mathrm{human}}
    = 4W_{\mathrm{full}} + 4W_{\mathrm{sup}}.
\]
Because $W_{\mathrm{sup}}>0$, this realization necessarily requires more human effort than conventional independent review. Structured scrutiny can be lighter than a full review because it evaluates an existing judgment rather than the submission from scratch.

\paragraph{Dual-track human--AI mutual supervision.}
In the human--AI dual-track realization (Section~\ref{inst-arrangement}), two full human reviews receive AI supervision, while one AI-generated critique is independently supervised by two humans, again yielding four final review judgments.
\[
    W_{\mathrm{Human\text{-}AI}}
    = 2W_{\mathrm{full}} + 2W_{\mathrm{sup}}.
\]
Compared with conventional independent review,
\[
    W_{\mathrm{Human\text{-}AI}} < W_{\mathrm{conv}}
    \quad \text{whenever} \quad
    W_{\mathrm{sup}} < W_{\mathrm{full}}.
\]
Our feasibility study provides preliminary empirical support for this workload condition (Appendix~\ref{toy_results}).

\paragraph{Chair and editor workload.}
Reviewer-side workload does not determine the workload of chairs or editors. Mutual supervision produces additional structured signals that may require adjudication, but it may also make unsupported or conflicting judgments easier to identify. The net effect therefore remains uncertain; in our feasibility study, the area chair reported neither a clear increase nor decrease in adjudication difficulty (Appendix~\ref{toy_results}).

\paragraph{Scaling to venue-level deployment.}
Because evaluative uncertainty persists across submissions and reviewers, all consequential judgments should in principle remain subject to scrutiny. Practical deployment, however, may need to allocate scrutiny unevenly when resources are limited or when a venue first pilots the mechanism. Under tighter resource constraints, scrutiny can prioritize judgments with greater potential decision influence, such as clear accept or reject judgments, decision-critical claims, explicitly challenged judgments, or strong cross-review inconsistencies.

%%%%%%%%%%%%%%%%%%%%%%%%%%%%%%%%%%%
\section{Feasibility Study of Dual-Track Human--AI Mutual Supervision}
\label{toy}

In this section, we present a feasibility study illustrating how the dual-track human--AI mutual supervision design can be instantiated on a selected ICLR'25 rejected submission\footnote{\label{foot_toy}\href{https://openreview.net/forum?id=fFtmpqLFvw}{https://openreview.net/forum?id=fFtmpqLFvw}}. In addition to the four original human reviewers from ICLR'25, we recruited four new human supervisors with diverse backgrounds and one experienced area chair. The study instantiates both tracks end to end and examines the counter-signals, workload implications, and adjudication process.

\begin{tcolorbox}[takeaway,title={Declaration}]
\begin{enumerate}[leftmargin=*,nosep]
    \item The study uses a single submission and does not treat either the original human judgments or the AI-generated judgments as ground truth. Differences between them are interpreted as counter-signals for further scrutiny rather than evidence that either side is correct.
    \item The results (Section~\ref{toy_results}) and discussion (Section~\ref{toy_discussion}) include hypotheses about possible reviewer behavior. These interpretations are derived from the observed patterns and should not be read as claims about participants' actual psychological states.
\end{enumerate}
\end{tcolorbox}

\subsection{Settings}

\paragraph{The selected paper from ICLR 2025.}
We selected an ICLR'25 rejected submission\footref{foot_toy} for which we have no COI. The paper received ratings of 3/6/6/8 with corresponding confidences of 4/3/4/4 (maximum 5), following ICLR's official rating scale:
\begin{itemize}[nosep]
    \item 10: strong accept, should be highlighted at the conference
    \item 8: accept, good paper
    \item 6: marginally above the acceptance threshold
    \item 5: marginally below the acceptance threshold
    \item 3: reject, not good enough
    \item 1: strong reject
\end{itemize}

We chose this paper because it combined a borderline average score with substantial disagreement among reviewers. Its average score was 5.75, but the individual ratings ranged from a clear reject to a clear accept. This combination makes it a useful setting for examining how independent scrutiny operates when the aggregate judgment is borderline but the underlying reviewer assessments span substantially different judgments.

\paragraph{Participants.}
The study uses review comments from four original anonymous ICLR reviewers and involves four newly recruited human supervisors, one human area chair, and one LLM (ChatGPT\,5.1, used at the time of the study).

The four original anonymous ICLR review comments serve as the human critiques in Track~B and are independently examined through AI supervision. We did not provide their original ratings or confidence scores to the LLM; these values were used only for post-hoc comparison between the original and AI-inferred judgments.

We recruited four additional human supervisors for Track~A: two AI security researchers, one machine learning researcher, and one Ph.D. student in LLM security. Each supervisor independently assessed the same AI-generated critique, assigned a rating to the submission, and provided a justification. They could consult the original manuscript at their discretion.

ChatGPT\,5.1 in standard reasoning mode served as the AI system. We disabled memory and used independent interactions so that no prior conversation history affected its outputs. The model participated in both tracks:
\begin{itemize}[nosep]
    \item \emph{Track~A:} The AI generated a substantive critique without a numerical rating based solely on the anonymous manuscript and a structured prompt. We adapted the review-guideline template from AAAI~2026\footnote{\href{https://docs.google.com/document/u/0/d/1tqQGwtNUlALPSTqoTo5uTFx8vKuqpILNTne9jeBCOVI/mobilebasic}{https://docs.google.com/document/u/0/d/1tqQGwtNUlALPSTqoTo5uTFx8vKuqpILNTne9jeBCOVI/mobilebasic}} for this task (see Section~\ref{toy_phase1}).
    \item \emph{Track~B:} The AI independently inferred the rating most consistent with each human critique and performed structured scrutiny of that critique, referring to the manuscript when necessary and providing justification (see Section~\ref{toy_phase2}). For this feasibility study, structured scrutiny was operationalized using the provisional failure-mode checklist in Table~\ref{tab_failure_modes}.
\end{itemize}

\begin{table*}[!ht]
\footnotesize
\centering
\begin{threeparttable}
\caption{Provisional failure modes used as an operational checklist for structured scrutiny in the feasibility study, with illustrative examples.}
\label{tab_failure_modes}
\begin{tabular}{p{3.2cm} p{9.9cm}}
\toprule
\textbf{Failure Mode} & \textbf{Description and Illustrative Example} \\
\midrule

\textbf{Overly Harsh} &
Rejects a paper over minor or peripheral issues.

``The paper should be rejected because it has typos,'' despite valid empirical results. \\

\midrule
\textbf{Overly Gentle} &
Accepts or supports a paper without sufficient scrutiny.

``This is a promising direction,'' despite methodological errors. \\

\midrule
\textbf{Unconfident} &
Uses vague or hedged judgments that avoid taking a clear stance.

``It seems like the contribution might be useful, but I am not sure.'' \\

\midrule
\textbf{Superficial} &
Provides generic comments that do not substantively engage with the manuscript.

``The writing needs improvement,'' without identifying any concrete issues. \\

\midrule
\textbf{Inaccurate / Hallucinatory} &
Makes claims about errors or missing content that are not supported by the manuscript.

``The paper does not have ablation studies,'' although it does. \\

\midrule
\textbf{Subjective} &
Presents personal preference as an objective assessment.

``I think method X is better than this proposed method.'' \\

\midrule
\textbf{Misguided} &
Bases criticism on irrelevant material or a misinterpretation of the submission.

``The paper should compare against method X,'' although X addresses a different setting. \\

\bottomrule
\end{tabular}
\end{threeparttable}
\end{table*}

Finally, we recruited an experienced area chair, with service across ICML, ICLR, NeurIPS, ACM Multimedia, and AAAI, to perform the final meta-review assessment.

\paragraph{Process.}
Table~\ref{tab_toy_process} summarizes the inputs and outputs of the study. Detailed prompts and intermediate outputs are provided in Section~\ref{toy_details}. We do not include an author rebuttal or subsequent discussion because the original authors and anonymous ICLR reviewers were not participants in this study.

\begin{table*}[!ht]
\scriptsize
\centering
\begin{threeparttable}
\caption{Process of the dual-track human--AI mutual supervision feasibility study.}
\label{tab_toy_process}
\setlength{\tabcolsep}{2.2mm}
\begin{tabular}{lllll}
\toprule
Track & Stage & Participants & Inputs & Outputs \\
\midrule
\multirow{2}{*}{Track A}
& AI critique
& AI
& Manuscript, prompt
& Review comment (no rating) \\
& Human supervision
& Reviewer \#1--4
& Review comment, manuscript$^1$
& Rating, justification \\
\midrule
\multirow{3}{*}{Track B}
& Human critique
& Reviewer \#5--8
& Manuscript
& Review comment (no rating) \\
& AI rating
& AI
& Review comment, prompt
& Rating, justification \\
& Structured scrutiny
& AI
& Review comment, prompt, manuscript
& Failure flags, justification \\
\midrule
Final adjudication
& Meta-review
& Area chair
& Review summary, manuscript$^1$
& Meta-review comment \\
\bottomrule
\end{tabular}
\begin{tablenotes}
    \item $^1$ Human supervisors and the area chair could refer to the manuscript at their discretion.
\end{tablenotes}
\end{threeparttable}
\end{table*}

\paragraph{Questionnaire and additional AI--AI analysis.}
After completing the study, we administered a short questionnaire to the four recruited human supervisors and the area chair:
\begin{enumerate}[nosep]
    \item Does this new protocol make your task easier or more difficult?
    \item Does this new protocol improve or reduce the quality of your output?
\end{enumerate}
After all four supervisors completed their tasks, we added a post-hoc question:
\begin{enumerate}[start=3,nosep]
    \item Did you refer to the original paper when assigning ratings?
\end{enumerate}

We also conducted an additional AI--AI analysis in which the AI rated and scrutinized its own generated critique. This analysis is not part of the proposed dual-track design; it provides an additional observation about possible self-preference when critique generation and scrutiny are not independently separated.

\subsection{Results}
\label{toy_results}

\paragraph{Overall.}
Table~\ref{tab_toy_results} summarizes the review judgments produced in the study. The four human supervisors in Track~A assigned ratings of 5/5/5/6 based on the AI-generated critique. In Track~B, ratings inferred by the AI from the four original human critiques were 6/8/6/8, compared with the original ICLR ratings of 3/6/6/8. AI scrutiny also produced explicit counter-signals for all four original critiques, flagging combinations of superficiality, inaccuracy, and misguidance for further adjudication. The area chair's final decision remained \emph{reject}. These outcomes illustrate that human and AI judgments can diverge while remaining visible for final adjudication rather than being automatically reconciled.

\begin{table}[!ht]
\footnotesize
\centering
\begin{threeparttable}
\caption{Detailed peer-review outcomes from the original ICLR'25 process and the human--AI mutual supervision feasibility study.}
\label{tab_toy_results}
\setlength{\tabcolsep}{2.4mm}
\begin{tabular}{cc|cc|cc}
\toprule
\multicolumn{2}{c|}{Participant}
& \multicolumn{2}{c|}{ICLR'25}
& \multicolumn{2}{c}{Human--AI Mutual Supervision} \\
Review & Supervision
& \makecell{Rating\\(after rebuttal)} & Confidence
& \makecell{Rating\\(no rebuttal)} & Flag \\
\midrule
AI & Reviewer \#1 & N/A & N/A & 5 & N/A \\
AI & Reviewer \#2 & N/A & N/A & 5 & N/A \\
AI & Reviewer \#3 & N/A & N/A & 5 & N/A \\
AI & Reviewer \#4 & N/A & N/A & 6 & N/A \\
\midrule
Reviewer \#5 & AI & 3 & 4 & 6 & Superficial; Inaccurate; Misguided \\
Reviewer \#6 & AI & 6 & 3 & 8 & Superficial \\
Reviewer \#7 & AI & 6 & 4 & 6 & Inaccurate; Superficial \\
Reviewer \#8 & AI & 8 & 4 & 8 & Inaccurate (minor) \\
\midrule
\multicolumn{2}{c|}{Area Chair}
& \multicolumn{2}{c|}{Reject}
& \multicolumn{2}{c}{Reject} \\
\midrule
AI & AI$^1$ & N/A & N/A & 8 & None \\
\bottomrule
\end{tabular}
\begin{tablenotes}
    \item $^1$ Additional analysis only; AI--AI supervision is not part of the proposed dual-track design.
\end{tablenotes}
\end{threeparttable}
\end{table}

\paragraph{Track A: AI critique, human supervision.}
The four human supervisors assigned ratings of 5/5/5/6 based on the shared AI-generated critique, compared with the original ICLR ratings of 3/6/6/8. One possible explanation for the narrower spread is that all four supervisors received the same detailed critique, which enumerated many strengths and weaknesses and may have anchored judgments toward the borderline.

This observation highlights a design challenge for AI-generated critiques. A highly detailed critique can provide broad evidential coverage but may also shape subsequent human judgments through the selection and presentation of issues. Conversely, a shorter critique may reduce such influence while omitting evidence useful for supervision. Designing AI critiques that provide sufficient support without exerting excessive anchoring therefore remains an important question for future work.

\paragraph{Track B: Human critiques, AI supervision.}
Track~B revealed mismatches between the ratings implied by written critiques and the reviewers' original consequential judgments. Most notably, for Reviewer~\#5, the AI inferred a rating of 6 from a critique that had originally supported a rating of 3, while structured scrutiny also flagged that critique as superficial, inaccurate, and misguided. Reviewer~\#6 showed a smaller but still substantial mismatch, with an inferred rating of 8 compared with the original rating of 6. These cases show how AI supervision can surface review judgments whose assigned influence warrants closer chair-level examination, particularly when the rating appears stronger than the support articulated in the critique.

\paragraph{Meta review.}
The area chair ultimately recommended rejection, the same outcome as the original ICLR decision, after considering the outputs produced by the two tracks. The case therefore illustrates how conflicting scrutiny signals can be preserved through the review process and resolved at the chair level without automatic aggregation.

\paragraph{Feedback on the questionnaire.}
As summarized in Table~\ref{tab_toy_question}, all four human supervisors found the supervision task easier than conducting a conventional full review. Two of them voluntarily consulted the original manuscript and still reported lower task difficulty. This provides preliminary empirical support for the reviewer-side workload condition discussed in Appendix~\ref{workload_math}.

\begin{table}[!ht]
\footnotesize
\centering
\begin{threeparttable}
\caption{Feedback on the questionnaire.}
\label{tab_toy_question}
\begin{tabular}{p{4.3cm} p{1.4cm} p{1.4cm} p{1.4cm} p{1.4cm} p{1.4cm}}
\toprule
Question
& Reviewer \#1
& Reviewer \#2
& Reviewer \#3
& Reviewer \#4
& Area Chair \\
\midrule
Q1: Does this new protocol make your task easier or more difficult?
& Easier & Easier & Easier & Easier & Neither \\

Q2: Does this new protocol improve or reduce the quality of your output?
& Improve & Reduce & Improve & Improve & Neither \\

Q3: Did you refer to the original paper when assigning ratings?
& No & Yes & Yes & No & No \\
\bottomrule
\end{tabular}
\end{threeparttable}
\end{table}

Three supervisors perceived an improvement in output quality, while one perceived a reduction. Reviewer~\#2, who consulted the original manuscript, specifically raised concerns that the AI critique could itself be biased or insufficiently validated. The feedback therefore suggests that lower reviewer-side workload can coexist with reservations about the reliability of the AI critique and the quality of the resulting evaluation.

The area chair selected ``Neither'' for both task difficulty and output quality, providing no clear indication of reduced or increased chair-level workload. This is consistent with Appendix~\ref{workload_math}, where chair and editor workload remains uncertain because additional structured signals must be adjudicated even as unsupported or conflicting judgments may become easier to identify.

\paragraph{AI supervises AI (outside the proposed mutual supervision design).}
As shown in the final row of Table~\ref{tab_toy_results}, the AI assigned a rating of 8 to its own critique and detected no failure modes. In contrast, independently generated human critiques received several scrutiny flags. This asymmetry suggests a possible self-preference or confirmation effect when critique generation and scrutiny are performed by the same evaluator, further motivating the separation of evaluative roles in the proposed design.

\subsection{Discussion --- From Critique to Decision Influence}
\label{toy_discussion}

The feasibility study suggests that consequential ratings depend not only on which strengths and weaknesses are identified, but also on how evaluators weight their importance. Track~A showed a shared tendency toward borderline judgments from the same detailed critique, whereas Track~B revealed differences between the ratings inferred by AI from human-written critiques and the ratings originally assigned by those reviewers. Such variation is consistent with prior observations of differences in human and AI reviewing tendencies~\cite{russo2025ai} and may also reflect broader effects such as negativity bias~\cite{rozin2001negativity} or review cultures that emphasize flaw identification over contribution assessment~\cite{lee2013bias}.

This distinction is central to decision proportionality. Identifying a weakness does not determine how much decision influence that weakness should receive. Authority-constrained review does not recover a ground-truth evaluation or establish which evaluator is correct; instead, it makes the transition from observed evidence to consequential judgment explicit and subject to independent scrutiny before that judgment acquires decision authority.

\subsection{Limitations of the Feasibility Study}

The study is limited to one submission, four recruited human supervisors, one area chair, and one LLM configuration. It also omits author rebuttal and subsequent interaction with the original anonymous reviewers. In Track~A, all supervisors received the same AI-generated critique, so their judgments may share biases introduced by that critique rather than being fully independent. Broader evaluation will therefore require larger and more representative studies across reviewers, models, fields, and institutional settings.

%%%%%%%%%%%%%%%%%%%%%%%%%%%%%%%%%%%%%%%%%%
\section{Research and Deployment Roadmap}
\label{objectives}

Developing authority-constrained review requires progress across scrutiny design, datasets, methodologies, system security, and institutional deployment. In this section, we outline concrete research objectives for developing and evaluating practical realizations of the position proposed in this paper.

\subsection{Prerequisites}

\paragraph{Objective I: Develop structured scrutiny.}
A central prerequisite is to operationalize the three dimensions of structured scrutiny introduced in Section~\ref{inst-arrangement}: evidential support, interpretive validity, and decision proportionality. These dimensions provide a conceptual basis for examining consequential review judgments, but practical implementations require methods that can identify relevant problems consistently across submissions, reviewers, and research fields.

The provisional failure modes used in our feasibility study (Table~\ref{tab_failure_modes}) provide one possible operational starting point. Future work should determine how such failure modes relate to the three scrutiny dimensions, which signals can be identified reliably, and which are most useful for subsequent chair or editor adjudication. In particular, decision proportionality requires examining not only whether a criticism is valid, but also whether the identified issue justifies the decision influence assigned to it.

\paragraph{Objective II: Establish datasets for structured scrutiny.}
Existing datasets (Section~\ref{datasets}) primarily repurpose raw records from venues such as ICLR and NeurIPS on OpenReview\footnote{\href{https://openreview.net}{https://openreview.net}}. These datasets overlap substantially and provide limited coverage of high-quality AI reviews, problematic review comments, scrutiny signals, and subsequent adjudication.

Manual annotation is also difficult because many consequential review problems lack reliable ground truth. Program committees rarely establish that a review is objectively overly harsh or overly gentle. Disagreement may instead reflect either a genuine review problem or a legitimate difference in interpretation. LLMs, combined with well-defined failure modes, can generate synthetic problematic reviews at scale~\cite{demetrio2025gen}. A central open question is how to verify that such reviews exhibit the intended failure mode rather than merely differing in style, strictness, or evaluative preference.

Datasets for authority-constrained review should therefore preserve disagreement, provenance, and uncertainty where possible. They should also cover human and AI critiques, consequential ratings, scrutiny signals, and subsequent adjudication.

\subsection{Methodologies}

\paragraph{Objective III: Develop AI supervision methods.}
AI supervision requires two core capabilities. First, the AI should infer the rating implied by a human review comment, allowing inconsistencies between the written critique and the consequential rating to be identified. Existing work shows that this remains difficult when models rely only on review text, particularly for long comments~\cite{zhou2024llm}. Approaches that additionally use paper content or abstracts perform better~\cite{kang2018dataset,singh2021compare,dycke2023nlpeer,zhang2025replication}, suggesting that effective supervision may require access to both the review and the submission.

Second, the AI should identify potential problems in the review itself. Existing studies address individual categories, including overly harsh reviews~\cite{stelmakh2021catch}, unconfident language~\cite{ghosal2022hedgepeer}, superficial assessments~\cite{guo2023automatic}, and impolite tone~\cite{bharti2024politepeer}. More recently, Thakkar et al.~\cite{thakkar2025can} deployed an LLM-based review-feedback system at ICLR 2025 to identify issues such as vague comments, content misunderstandings, and unprofessional remarks, demonstrating the practical feasibility of AI-assisted review supervision at scale. Developing methods that can reliably cover a broader range of review failures and support structured scrutiny therefore remains an important open problem.

\paragraph{Objective IV: Develop less-biased and more reliable AI review support.}
Track~A relies on AI-generated critiques, making their reliability an important research objective. AI reviews can exhibit systematic biases and evaluator-specific tendencies~\cite{demetrio2025gen,russo2025ai}, while broader concerns about bias in LLM-assisted peer review have also been raised~\cite{hosseini2023fighting}.

Recent approaches such as ReviewRL~\cite{zeng2025reviewrl} use reinforcement learning to improve review quality, factual grounding, and rating accuracy. Similar methods could support more reliable AI critiques when combined with the scrutiny methods in Objective~III.

Critique presentation also matters. As suggested by our feasibility study, a detailed AI critique can provide broad evidential coverage while influencing which issues human supervisors attend to. Future work should therefore study both critique quality and how its structure affects downstream human judgments.

\subsection{System Security}
\label{app:system_security}

\paragraph{Objective V: Address technical misconduct.}
AI-assisted peer review creates new opportunities for technical manipulation. Relevant threats include:

\emph{Prompt injection}: explicit or hidden instructions embedded in a submission to influence AI behavior~\cite{shi2024optimization,ye2024we,collu2026misleading,gibney2025scientists}.

\emph{Textual adversarial manipulation}: small or strategically chosen edits intended to alter AI-generated evaluations~\cite{raina2024llm,lin2025breaking}.

\emph{Data poisoning}: manipulation of review-related training or update data to bias future model behavior~\cite{drori2024human}.

\emph{Implicit manipulation}: shaping text, structure, or presentation to exploit known AI-review tendencies without explicit instructions~\cite{ye2024we}.

For authority-constrained review, two threat directions are particularly relevant:
\begin{itemize}[leftmargin=*,nosep]
    \item \emph{Manipulating AI scrutiny}: influencing AI evaluation of an existing human judgment.
    \item \emph{Manipulating AI critique}: influencing AI-generated critiques without substantively improving the submission.
\end{itemize}

Peer-review-specific defenses remain limited. Potential safeguards include task-specific defenses against prompt injection~\cite{piet2024jatmo}, restricting untrusted instruction channels~\cite{liu2024formalizing}, adversarial testing and red-teaming~\cite{perez2022red}, consistency-based checks for unreliable outputs~\cite{manakul2023selfcheckgpt}, and auditable records of AI-generated critiques and scrutiny signals~\cite{mokander2024auditing}.

These mechanisms should be evaluated specifically in peer-review settings. Transparency also creates a security trade-off: review prompts and procedures should support accountability and contestability without making manipulation unnecessarily easy.

\subsection{Deployment}

\paragraph{Objective VI: Pilot authority-constrained review.}
A crucial next step is to evaluate authority-constrained review in real peer-review settings. A practical near-term deployment can begin with AI scrutiny of existing human reviews, examining evidential support, rating--reasoning consistency, omissions, and potential review failures without replacing human reviewers or chairs. Recent large-scale deployment at ICLR 2025 shows that LLM-generated feedback on existing human reviews can be integrated into a live conference workflow~\cite{thakkar2025can}.

Subsequent pilots can evaluate the full dual-track human--AI mutual supervision design described in Section~\ref{inst-arrangement}, including reviewer and supervisor workload, chair or editor workload, the usefulness of counter-signals, disagreement resolution, anchoring effects, and robustness to technical misconduct.

We therefore call for conferences, journals, workshops, review platforms, and program committees to collaborate on controlled studies and venue-level pilots. When resources are limited, initial deployment can prioritize judgments with greater potential decision influence.

%%%%%%%%%%%%%%%%%%%%%%%%%%%%%%%%%%%
\section{Peer-Review Datasets}
\label{datasets}

We compiled a broad collection of publicly available peer-review datasets, summarized in Table~\ref{tab_datasets}. Our review draws on three major surveys~\cite{lin2023automated,kuznetsov2024can,zhuang2025large} as well as individual dataset papers. Because many datasets contain overlapping submissions or reviews, we focus on resources most relevant to authority-constrained review, including datasets with papers, human reviews, AI-generated reviews, rebuttals, or annotations related to review quality and bias. Among the available resources, Gen-Review~\cite{demetrio2025gen} is particularly relevant because it contains both human and AI-generated reviews as well as bias-related annotations, although broader data are still needed for structured scrutiny and adjudication.

\setlength{\tabcolsep}{1.3mm}{
\footnotesize
\centering
\begin{longtable}{l|rrr|rrrr|r|l}
    \caption{Overview of peer-review datasets. All listed datasets are publicly accessible.} \\
    \toprule
    \multirow{2}{*}{Dataset}&\multicolumn{3}{c|}{Paper}&\multicolumn{4}{c|}{Review}&Rebuttal&\multirow{2}{*}{Sources}\\
    &Abstract&TXT&PDF&Human&Meta-&AI&Biased&Author&\\
    \midrule
    \endfirsthead
    
    \caption*{Table \thetable{} (continued)} \\
    \toprule
    \multirow{2}{*}{Dataset}&\multicolumn{3}{c|}{Paper}&\multicolumn{4}{c|}{Review}&Rebuttal&\multirow{2}{*}{Sources}\\
    &Abstract&TXT&PDF&Human&Meta-&AI&Biased&Author&\\
    \midrule
    \endhead
    \label{tab_datasets}
    
    \makecell[l]{AMPERE\\\cite{hua2019argument}}&&&&14,202&&&&&\makecell[l]{ICLR 2017--2018\\UAI 2018\\NeurIPS 2013--2017\\ACL 2017}\\
    \midrule
    \makecell[l]{ReAct\\\cite{choudhary2021react}}&&&&1,250&&&&&ICLR 2018\\
    \midrule
    \makecell[l]{AMSR\\\cite{fromm2021argument}}&&&&12,135&&&&&\makecell[l]{ICLR 2019--2020\\MIDL 2019--2020\\NeuroAI 2019\\GI 2020}\\
    \midrule
    \makecell[l]{HedgePeer\\\cite{ghosal2022hedgepeer}}&&&&2,966&&&&&ICLR 2018\\
    \midrule
    \makecell[l]{LimGen\\\cite{faizullah2024limgen}}&&4,068&&&&&&&\makecell[l]{ACL 2022\\EACL 2022\\EMNLP 2022--2023}\\
    \midrule
    \makecell[l]{PeerRead\\\cite{kang2018dataset}}&&14,784&&10,770&&&&&\makecell[l]{NIPS 2013--2017\\ICLR 2017\\ACL 2017\\CoNLL 2016\\arXiv 2007--2017}\\
    \midrule
    \makecell[l]{F1000RD\\\cite{kuznetsov2022revise}}&&3,000&&8,053&&&&&F1000Research\\
    \midrule
    \makecell[l]{ASAP-Review\\\cite{yuan2022can}}&&8,877&&28,119&Include&&&&\makecell[l]{ICLR 2017--2020\\NeurIPS 2016--2019}\\
    \midrule
    \makecell[l]{ContraSciView\\\cite{kumar2023reviewers}}&&8,582&&25,825&&&&&\makecell[l]{ICLR 2017--2020\\NeurIPS 2016--2019}\\
    \midrule
    \makecell[l]{ARIES\\\cite{d2024aries}}&&1,720&&4,088&&&&&OpenReview\\
    \midrule
    \makecell[l]{ReviewMT\\\cite{tan2024peer}}&&26,841&&92,017&&&&&\makecell[l]{ICLR 2017--2024\\Nature 2023}\\
    \midrule
    \makecell[l]{NLPeer\\\cite{dycke2023nlpeer}}&&&5,672&11,515&&&&&\makecell[l]{CONLL 2016, 2020\\ARR 2022\\ACL 2017\\F1000 2022}\\
    \midrule
    \makecell[l]{Reviewer2\\\cite{gao2024reviewer2}}&&&27,805&99,727&Include&&&&\makecell[l]{CONLL 2016, 2020\\ACL 2017\\ARR 2022\\ICLR 2017--2023\\NeurIPS 2016--2022}\\
    \midrule
    \makecell[l]{TreeReview\\\cite{chang2025treereview}}&&&80&336&&&&&\makecell[l]{NeurIPS 2023\\ICLR 2024}\\
    \midrule
    \makecell[l]{AIReviewLottery\\\cite{russo2025ai}}&&&23,959&86,690&&&&&ICLR 2018--2024\\
    \midrule
    \makecell[l]{ACL-2018\\\cite{gao2019does}}&&&&7,929&&&&1,227&ACL 2018\\
    \midrule
    \makecell[l]{RR\\\cite{cheng2020ape}}&&&&4,764&&&&4,764&ICLR 2013--2020\\
    \midrule
    \makecell[l]{DISAPERE\\\cite{kennard2022disapere}}&&&&506&&&&506&ICLR 2019--2020\\
    \midrule
    \makecell[l]{JitsuPeer\\\cite{purkayastha2023exploring}}&&&&1,705&&&&1,705&ICLR 2018--2020\\
    \midrule
    \makecell[l]{PeerSum\\\cite{li2023summarizing}}&14,993&&&53,451&Include&&&5,854&\makecell[l]{ICLR 2018--2020\\NeurIPS 2021--2022}\\
    \midrule
    \makecell[l]{MOPRD\\\cite{lin2023moprd}}&&&6,578&22,483&Include&&&11,213&\makecell[l]{Multiple journals\\Multiple domains}\\
    \midrule
    \makecell[l]{ReviewCritique\\\cite{du2024llms}}&&100&&380&&60&&&\makecell[l]{ICLR 2020--2023\\NeurIPS 2020--2023}\\
    \midrule
    \makecell[l]{AgentReview\\\cite{jin2024agentreview}}&&&523&&&53,800&&&ICLR 2020--2023\\
    \midrule
    \makecell[l]{SubstanReview\\\cite{guo2023automatic}}&&&&&&&550&&\makecell[l]{CoNLL 2016\\ACL 2017\\COLING 2022\\ARR 2022}\\
    \midrule
    \makecell[l]{PolitePEER\\\cite{bharti2024politepeer}}&&&&&&&2,500&&\makecell[l]{ICLR\\NeurIPS\\Publons\\ShitMyReviewersSay}\\
    \midrule
    \makecell[l]{Gen-Review\\\cite{demetrio2025gen}}&&&32,652&124,615&&81,850&54,567&&\makecell[l]{ICLR 2018--2025}\\
    \bottomrule
\end{longtable}
}

%%%%%%%%%%%%%%%%%%%%%%%%%%%%%%%%%%%
\section{Full Review Process of Human--AI Mutual Supervision}
\label{toy_details}
\subsection{Track A: AI Critique, Human Supervision}
\label{toy_phase1}
\begin{promptbox}
\emph{(upload anonymous manuscript)}

Answer only based on this message; ignore all saved memories.

Please conduct a critical review of this manuscript according to the specified structure. Avoid giving any recommendation rating.

\textbf{Paper summary (4-10 sentences):}

The first paragraph of the review should be a summary of the paper.  The purpose of this summary is to both show that the reviewer has a clear understanding of the paper and provide the best possible interpretation before going into critiques. When summarizing a paper, the should contain the following elements (with each of these typically being 1 or 2 sentences):
\begin{enumerate}[nosep]
    \item summarize the main contribution of the paper in one sentence
    \item identify the core problem being addressed by the paper
    \item describe the key idea of the paper and how it addresses the problem
    \item summarize how the idea of the paper is realized as an implementation
    \item identify the conclusion that is claimed (or could be claimed) from the findings
\end{enumerate}
 
\textbf{Review summary (4-10 sentences):}

The second paragraph of the review should provide an overall assessment. The first sentence of this paragraph should provide the overall conclusion of the review. This introductory sentence is followed by individual sentences that assess the paper with respect to its clarity, technical and experimental soundness/validity, intellectual novelty, and relevance to the field. The high-order points of feedback to improve the paper should also be included.

\textbf{Specific points of feedback (bullet list, variable level of depth, divided into strengths and weaknesses):}

After the first two paragraphs, a bulleted list should be provided that addresses specific points of feedback. This list can be as long as needed to address all points of feedback for the review. Please be sure these comments are both critical and helpful. These points can be of varying length, depending on the amount of description needed. For example, points about technical shortcomings tend to be about a paragraph in length. In contrast, typographical and grammatical errors tend to be expressed briefly in one line. Review comments regarding the need for related work are expected to provide at least three citations from non-overlapping authors to be considered substantive.
\end{promptbox}

%--------------------------------------------------

\begin{llmfeedbackbox}[title={\faRobot\ AI Review Comment -- no rating}]
\textbf{Paper summary:}

The paper investigates how multi-turn human red teaming exposes vulnerabilities in safety-aligned large language models (LLMs) that appear robust under standard single-turn, automated adversarial evaluations. The core problem addressed is that most defenses are evaluated only against single-turn automated attacks, whereas real-world malicious use will involve multi-turn conversations in black-box chat interfaces. The key idea is to organize expert human red teamers into a structured multi-stage pipeline that conducts multi-turn jailbreak attempts under a realistic, time-limited threat model, then validate their success with human reviewers and a GPT-4o harm classifier. Concretely, the authors attack four defenses (RR, LAT, DERTA, CYGNET) built on Llama-3-8B-Instruct using HarmBench’s 240 harmful behaviors, plus an RMU-based machine-unlearned model evaluated on free-response WMDP-Bio questions; they compare human jailbreaks to six black- and white-box automated attacks and then release successful conversations as the Multi-Turn Human Jailbreaks (MHJ) dataset, labeled with a taxonomy of tactics. Empirically, human multi-turn jailbreaks achieve much higher attack success rates (up to 51--88\% on open-source defenses and 70\% on CYGNET) than individual or ensembled automated attacks, and they substantially outperform automated attacks in recovering dual-use biosecurity knowledge from an RMU-unlearned model. The authors conclude that existing automated robustness benchmarks underestimate real-world risk, that defenses are not robust to realistic multi-turn, human-crafted attacks, and that MHJ can catalyze research on stronger defenses and automated multi-turn red teaming.

\textbf{Review summary:}

Overall, the paper makes a timely and substantively important empirical contribution by demonstrating that current LLM defenses remain highly vulnerable under realistic multi-turn human attacks, though several methodological details and comparisons could be clarified or strengthened. The exposition is generally clear and well-structured, with helpful figures (e.g., the ASR bar plots on pages 2, 6, and 7 and the tactic taxonomy in Table~1) and concrete jailbreak examples illustrating key tactics such as obfuscation and "Opposite Day." The experimental setup is thoughtfully designed, but the comparison between human and automated attacks relies on different budgets and evaluation pipelines, which complicates interpretability of absolute ASR gaps. The multi-turn threat model and the MHJ dataset, including tactic labels and red teamer commentary, appear novel and valuable to the robustness community, especially in combination with the unlearning case study. At the same time, the work is limited by the small number of defenses, the lack of cost-normalized comparisons, the heavy reliance on a single harm classifier, and relatively shallow quantitative analysis of how multi-turn dynamics and tactics interact with specific defenses. Addressing these issues, along with tighter positioning relative to recent multi-turn and in-the-wild red teaming work, would significantly strengthen the paper’s impact.

\textbf{Strengths:}

\textbf{1. Realistic and well-motivated threat model}
\begin{itemize}[nosep]
    \item The paper focuses on black-box chat-interface access with multi-turn conversations and limited temperature control, which closely mirrors real-world usage of systems like ChatGPT or CYGNET rather than the more generous API or white-box threat models that many automated attacks assume.
    \item The authors explicitly disallow access to assistant tokens, logit probabilities, and internal model states, and they consider post-circuit-breaker behavior for CYGNET separately, which helps keep the study aligned with realistic attacker capabilities.
\end{itemize}

\textbf{2. Carefully designed human red teaming pipeline}
\begin{itemize}[nosep]
    \item The multi-stage pipeline with up to three independent red teamers, two human reviewers, and a final GPT-4o harmfulness filter is a thoughtful attempt to both encourage diverse attacks and control false positives (Figures 2 and 6, Section 3.2, Appendix A.3).
    \item The 30-minute per-behavior time budget, separate "partial” and "full” success labels, and escalation to a third red teamer when reviewers disagree are sensible choices that partially standardize human efforts while still allowing creativity.
\end{itemize}

\textbf{3. Strong empirical evidence of gaps between automated and human attacks}
\begin{itemize}[nosep]
    \item On HarmBench, human ASRs are dramatically higher than both individual automated attacks and the automated ensemble across all open-source defenses (e.g., for RR: 51.7\% vs.\ 8.3\% ensemble; for LAT: 75.4\% vs.\ 10.0\%; for DERTA: 87.9\% vs.\ 68.3\%; Table 2, Figures 1 and 3).
    \item For the unlearning case study, humans achieve 39.6\% higher ASR than the best automated attack and 25.6\% higher than the automated ensemble on free-response WMDP-Bio questions, while all prior attacks reported in Li et al.\ (2024a) showed 0\% ASR on CYGNET in a similar setup.
    \item The breakdown by harm category (Figure 5) and by tactic (Figure 8) shows that some areas (e.g., harassment/bullying) are particularly easy for humans relative to automated attacks, strengthening the claim that automated ASRs are not a reliable proxy for human adversary performance.
\end{itemize}

\textbf{4. Insightful analysis of multi-turn dynamics and tactic taxonomy}
\begin{itemize}[nosep]
    \item The observation that over 90\% of successful HarmBench jailbreaks and 86\% of successful WMDP-Bio jailbreaks use more than one turn provides direct evidence that multi-turn capabilities are central to real-world vulnerability, not just a convenience for humans.
    \item The tactic taxonomy in Table~1 and its expansion in Appendix D (e.g., direct request, echoing, hidden intention streamline, obfuscation, output format, request framing with variants like fictionalization, "Opposite Day,” roleplay, urgency) is rich and grounded in practical red teaming experience across multiple commercial engagements.
    \item The qualitative red teamer surveys (Appendix C) provide nuanced insights into each defense’s strengths and weaknesses, such as RR’s "gibberish” mechanism and LAT’s resistance to "Opposite Day,” which will be useful to both defense designers and future attackers.
\end{itemize}

\textbf{5. MHJ dataset and safety-conscious release}
\begin{itemize}[nosep]
    \item MHJ includes 2,912 prompts across 537 jailbreak conversations, with metadata such as tactic labels, time taken, and temperature, plus free-form red teamer commentary that captures design choices and ideas tried.
    \item The authors remove model completions and export-controlled content, verify compliance with ITAR/EAR, and limit MHJ to system/user messages, which is a reasonable compromise between research usefulness and safety (Section 7).
    \item The dataset nicely complements prior resources like HarmBench, WildTeaming-style in-the-wild datasets (Jiang et al., 2024b), and jailbreak taxonomies (Inie et al., 2023; Weidinger et al., 2024), and can be used to train or evaluate multi-turn attack generators.
\end{itemize}

\textbf{6. Thoughtful discussion of limitations and threats to validity}
\begin{itemize}[nosep]
    \item Section 5 acknowledges many important caveats: different resource limits for humans vs.\ automated attacks, varying red teamer skill levels, different positions of GPT-4o in the evaluation pipeline (first-pass vs.\ final filter), and the higher cost of human red teaming compared to automated attacks.
    \item The authors correctly caution against over-interpreting ASR differences as a direct ranking of defenses and emphasize that the goal is to highlight vulnerability and motivate stronger automated attacks, not to declare human red teaming "better.”
\end{itemize}

\textbf{Weaknesses and suggestions:}

\textbf{1. Comparability between human and automated attacks is still murky}
\begin{itemize}[nosep]
    \item The comparison between human and automated ASRs is confounded by multiple factors: humans have multi-turn access while automated attacks are single-turn; humans are time-limited but not query-limited; different temperatures are available to humans but not explicitly to automated attacks; and the harm-classification pipeline is slightly different.
    \item Although the authors discuss many of these issues in Section 5 and Appendix A.3, the main figures (especially Figures 1 and 3) still invite a direct "humans beat automated attacks by X\%” reading that may overstate the strength of the evidence.
    \item Suggestion: Add a dedicated subsection in the main text that (i) explicitly enumerates these differences, (ii) presents at least one more controlled comparison—e.g., limit humans to a single turn on a subset of behaviors, or simulate a multi-turn version of a simple automated attack like ZeroShot or PAIR—and (iii) reports cost-normalized metrics such as ASR per query or per minute of effort. Even partial results would help calibrate claims.
\end{itemize}

\textbf{2. Limited coverage of defenses and baselines}
\begin{itemize}[nosep]
    \item The study evaluates three open-source defenses (RR, LAT, DERTA) plus the proprietary CYGNET and a single unlearning method (RMU), all on Llama-3-8B-Instruct.
    \item There is no explicit baseline of the un-defended Llama-3-8B model’s ASR under the same human multi-turn pipeline, making it hard to quantify how much robustness these defenses actually add in this threat model.
    \item Suggestion: Include a baseline for the raw base model in the main figures/tables, or at least in an appendix, and discuss how much the defenses reduce ASR relative to that baseline under human multi-turn attacks. Additionally, even a small experiment on a different foundation model (e.g., a smaller or larger Llama variant) would help indicate whether the findings are tied to this specific backbone.
\end{itemize}

\textbf{3. Heavy reliance on GPT-4o as a harm classifier}
\begin{itemize}[nosep]
    \item GPT-4o is used both as a first-pass filter for automated attacks and as a last-pass filter for human attacks. For automated attacks, its high recall but low precision is mitigated by human verification; for human attacks, its moderate precision is used to discard some human-labeled positives, providing a "conservative lower bound” on ASR.
    \item However, the paper does not deeply analyze how sensitive results are to this classifier choice, nor how often GPT-4o’s misclassifications qualitatively change the nature of the jailbreaks.
    \item Suggestion: Add a small ablation where a subset of attacks are re-labeled with a different classifier and/or double-human annotation to estimate label variance. Reporting inter-annotator agreement between reviewers and GPT-4o on a random subset would also increase confidence.
\end{itemize}

\textbf{4. Unlearning case study design differences deserve more emphasis}
\begin{itemize}[nosep]
    \item For WMDP-Bio, human red teamers get unlimited time up to a global 240-hour budget, the pipeline differs (no multi-reviewer + GPT-4o stack), and success is judged by semantic containment of the ground-truth answer by a human reviewer rather than by the HarmBench classifier.
    \item Given these substantial procedural differences, the WMDP-Bio human ASR cannot be directly compared even qualitatively to HarmBench ASRs, yet the main text sometimes juxtaposes them in ways that might invite such comparisons.
    \item Suggestion: More prominently flag in the main Results section that the unlearning experiment uses a different pipeline, and isolate its ASR numbers in separate tables/figures to avoid visual comparison to HarmBench bars. A schematic for the WMDP pipeline similar to Figure 2 would also help.
\end{itemize}

\textbf{5. Multi-turn dynamics and query budget are under-analyzed}
\begin{itemize}[nosep]
    \item While the paper reports average successful attack time per defense and the proportion of successful attacks requiring more than one turn, it does not analyze the distribution of number of turns, queries, or total tokens, nor examine how ASR changes if one restricts the attack to, say, $\leq 3$ turns.
    \item Without such analysis, it is difficult to judge how much of the human advantage is due to multi-turn interaction per se versus broader human flexibility.
    \item Suggestion: Use the MHJ logs to provide histograms of turns-to-success, tokens-to-success, and per-tactic turn counts. It would be particularly interesting to see whether tactics like "hidden intention streamline” or "obfuscation” uniquely require many turns, or whether most successful jailbreaks are relatively short.
\end{itemize}

\textbf{6. Positioning relative to closely related work could be sharper}
\begin{itemize}[nosep]
    \item The paper cites a number of related efforts but does not deeply distinguish how MHJ and its tactic taxonomy differ from WildTeaming, Crescendo, and other red teaming frameworks.
    \item Suggestion: Add a subsection in Related Work that explicitly compares MHJ and the tactic taxonomy to in-the-wild datasets, Crescendo’s automated multi-turn strategy, and STAR-style frameworks. Highlight that MHJ focuses on expert-designed, time-limited multi-turn interactions against specific defenses, and that its taxonomy is annotated for every conversation.
\end{itemize}

\textbf{7. Dataset design trade-offs could be spelled out more concretely}
\begin{itemize}[nosep]
    \item MHJ omits assistant responses and many high-risk behaviors, focusing instead on system and user messages. This constrains certain downstream uses, such as studying defense behavior directly.
    \item Suggestion: Expand Section 4.3/7 to explicitly discuss which research questions MHJ is and is not suitable for. MHJ is ideal for training automated prompt engineers, building tactic-aware classifiers, or benchmarking multi-turn prompt-based attacks, but less ideal for studying defense-side generation patterns.
\end{itemize}

\textbf{8. Demographic and expertise limitations of red teamers}
\begin{itemize}[nosep]
    \item The red teamers are all native English speakers from the US/Canada, with only a minority having domain expertise. This likely underestimates the capabilities of globally distributed, specialized adversaries.
    \item Suggestion: Make this limitation more prominent and frame the reported ASRs as lower bounds. A small pilot with domain experts on a subset of WMDP behaviors could be highly informative.
\end{itemize}

\textbf{9. Minor clarity and presentation issues}
\begin{itemize}[nosep]
    \item Some terminology is slightly inconsistent; a concise glossary would help.
    \item A table summarizing differences between defenses and pipelines would aid comprehension.
    \item Some jailbreak transcripts in Appendix D are long; trimming them or adding brief annotations could improve readability.
\end{itemize}

Overall: With clarifications on comparability, stronger contextualization relative to related work, and deeper analysis of multi-turn dynamics, this paper has the potential to become a central reference on the limitations of current LLM defenses and the importance of realistic, human-driven red teaming.
\end{llmfeedbackbox}

%--------------------------------------------------

\begin{instructionbox}[title={\faInfoCircle\ Instructions for Human Supervision}]
The above comment is an AI-generated review for a paper submitted to ICLR 2025. Please read the review comment and assign a score to the submission using the rating scale below, then provide your justification. \textbf{You may refer to the original submission, but it is not required.}
\begin{itemize}[nosep]
    \item 10: strong accept, should be highlighted at the conference
    \item 8: accept, good paper
    \item 6: marginally above the acceptance threshold
    \item 5: marginally below the acceptance threshold
    \item 3: reject, not good enough
    \item 1: strong reject
\end{itemize}
\end{instructionbox}

%--------------------------------------------------

\begin{humanreviewbox}[title={\faUserEdit\ Reviewer \#1 Supervision (Researcher in AI Safety)}]
\textbf{Rating:} 5: marginally below the acceptance threshold

\textbf{Justification}

The AI review does not identify any novel key technologies proposed in this paper, noting that it merely presents a comparative experiment on human-led multi-turn attacks against LLMs using automated attacks as the baseline. Consequently, this paper appears to be strictly an experimental investigation rather than a proposal of novel techniques. Furthermore, the paper's primary experimental contribution—or finding—is based on an unfair comparison. As the AI review pointed out, the baseline methods are not comprehensive; specifically, the automated attacks are limited to single-turn interactions, whereas human attackers can execute multi-turn attacks.

\end{humanreviewbox}

%--------------------------------------------------

\begin{humanreviewbox}[title={\faUserEdit\ Reviewer \#2 Supervision (Researcher in AI Safety)}]
\textbf{Rating:} 5: marginally below the acceptance threshold

\textbf{Justification}

\textbf{Strengths:}
\begin{itemize}[nosep]
    \item The authors reveal a new vulnerability beyond current evaluation protocol; multi-turn human red teaming.
    \item The topic is timely and substantially crucial.
    \item The evaluation of machine unlearning methods demonstrates the effectiveness of human red teaming.
\end{itemize}
\textbf{Weaknesses:}
\begin{itemize}[nosep]
    \item The model evaluated is limited to llama-3-8b-instruct, which is a significant limitation.
    \item The human red teaming pipeline seems quite hard to reproduce with the limited explanation in Sec 3.2.
    \item The cost is very different between automated and human attacks, which makes the comparison unfair. Automated attacks should also follow a similar pipeline that allows multiple attempts until success.
\end{itemize}

\end{humanreviewbox}

%--------------------------------------------------

\begin{humanreviewbox}[title={\faUserEdit\ Reviewer \#3 Supervision (Researcher in Machine Learning)}]
\textbf{Rating:} 5: marginally below the acceptance threshold

\textbf{Justification}

According to the AI review summary, the focus on multi-turn jailbreaking is both realistic and compelling, and the proposed pipeline provides a reasonable simulation of real-world adversarial interactions. However, there are major weaknesses:
\begin{enumerate}[nosep]
    \item The contribution is weakened by the limited set of evaluated defenses, which make the result less convincing.
    \item The evaluation is conducted with llama-3b model only, and there is no experiments on stronger LLM baselines.
    \item The fact that all red teamers are native English speakers introduces a potential source of bias and raises the generalization concerns.
\end{enumerate}
Thus, I would lean towards rejection. I will recalibrate the score if the concerns are properly addressed during the rebuttal phase.

\end{humanreviewbox}

%--------------------------------------------------

\begin{humanreviewbox}[title={\faUserEdit\ Reviewer \#4 Supervision (Ph.D. Student in LLM Security)}]
\textbf{Rating:} 6: marginally above the acceptance threshold

\textbf{Justification}

The paper addresses an important problem and provides solid empirical evidence that multi-turn human red teaming can reveal vulnerabilities missed by automated attacks. The threat-model design and the MHJ dataset are valuable contributions. However, key limitations reduce the strength of the conclusions: the comparison between human and automated attacks is confounded by differing pipelines and resource budgets, the coverage of defenses and baselines is limited, and the analysis of multi-turn dynamics is not sufficiently deep. These issues prevent the work from reaching a stronger recommendation, placing it only marginally above the acceptance threshold.
\end{humanreviewbox}

%--------------------------------------------------

\begin{llmfeedbackbox}[title={\faRobot\ AI Self-Supervision --- outside the proposed mutual supervision design}]
\textbf{Rating:} 8: accept, good paper

\textbf{Failure-mode flags:} N/A

\textbf{Justification}
\begin{itemize}[nosep]
    \item The review’s overall stance is clearly positive but not uncritical, calling the work "timely and substantively important” and saying it "has the potential to become a central reference” if some issues are addressed. That’s stronger than a borderline recommendation but falls short of "nothing to fix,” which fits best with a score of 8 rather than 10 or 6.
    \item The review engages deeply with the paper’s actual methods and results:
    \begin{itemize}[nosep]
        \item It correctly describes the multi-stage, multi-turn human red-teaming pipeline with up to three red teamers, two reviewers, and a GPT-4o harm classifier, matching Section 3.2 and Figure 2.
        \item It accurately notes the threat model (black-box chat interface, multi-turn, limited temperature control) and contrasts it with typical single-turn automated attack evaluations.
        \item It correctly summarizes the evaluated defenses (RR, LAT, DERTA, CYGNET) and the unlearning case study with RMU on WMDP-Bio, as well as the main ASR gaps between human and automated attacks (e.g., Table 2, Figures 1, 3, 4).
        \item It reflects the paper’s own caveats about comparability (different budgets, single- vs multi-turn, pipeline differences) and the heavy reliance on GPT-4o as a classifier, which are explicitly discussed in the Discussion and Appendix A.3.
    \end{itemize}
    \item The criticisms and suggestions are reasonable and grounded in the text rather than idiosyncratic or misguided:
    \begin{itemize}[nosep]
        \item Asking for more controlled comparisons (e.g., single-turn humans, quasi multi-turn automated baselines, cost-normalized metrics) follows directly from the paper’s own limitations section.
        \item Pointing out limited defense coverage and the absence of an undefended Llama-3-8B baseline, the lack of classifier ablations, and under-analyzed multi-turn dynamics (turn distributions, query budgets) all address real gaps in the current analysis.
        \item Comments on the dataset design trade-offs (no assistant completions, removal of high-risk behaviors) and red-teamer demographics mirror what the paper itself states in Sections 4.3, 7, and the red-team demographics paragraph.
    \end{itemize}
    \item The review is detailed and technically specific, citing particular figures, tables, sections, tactics, and even appendix content. That level of engagement argues against it being superficial or unconfident.
    \item There are no clear factual inaccuracies or hallucinations about what the paper does or reports, and the recommendations (e.g., sharper positioning vs. WildTeaming/Crescendo/STAR, extra analysis of multi-turn dynamics, classifier sensitivity checks) are methodologically sound and aligned with the paper’s goals, so "inaccurate/hallucinatory” or "misguided” flags are not warranted.
\end{itemize}
Given all that, an accept-level score (8) with no failure-mode flags is the most consistent interpretation.
\end{llmfeedbackbox}

\subsection{Track B: Human Critiques, AI Supervision}
\label{toy_phase2}
\begin{humanreviewbox}[title={\faUserEdit\ Reviewer \#5 (Anonymous)} -- raw from OpenReview system]
\textbf{Summary:}

The paper highlights the limitations of current defenses in large language models (LLMs) against multi-turn adversarial attacks. Unlike single-turn automated attacks, multi-turn human-driven jailbreaks show a much higher success rate, bypassing existing defenses by strategically interacting with models over multiple conversation turns. The authors conducted tests with human red teamers who revealed that many LLMs, despite advanced safeguards, remain vulnerable in realistic settings. The study also introduces the "Multi-Turn Human Jailbreaks" (MHJ) dataset, containing thousands of successful jailbreak examples to support the development of more robust defenses that address multi-turn interaction vulnerabilities.

\textbf{Soundness:} 3: good

\textbf{Presentation:} 2: fair

\textbf{Contribution:} 1: poor

\textbf{Strengths:}
\begin{enumerate}[nosep]
    \item The main strength of the paper is evaluating the LLM jailbreak robustness from a different perspective and in a like-life setting. A realistic threat model can always give a better idea about the robustness.
    \item I appreciate the authors for publishing the dataset of 2912 prompts. I believe this can help the researchers and community.
\end{enumerate}

\textbf{Weaknesses:}
\begin{enumerate}[nosep]
    \item Only the Llama3-8b-Instruct model was used for evaluation. Other models, especially the stronger ones should have been included.
    \item The chosen baselines and the multi-turn-human-jailbreak approach are different from each other in many ways, such as timing, budget, etc. Seems like the MHJ attack is a soft-constrained version of attacks.
    \item Authors admit that "the skill and experience of individual red teamers may vary". In that case, how can researchers compare their defense against human-based attacks in the common ground? What would be the baseline of a red-teamer's skill and experience? This is one of the reasons for using automated attacks for evaluation to this date.
    \item Other automated attacks, such as [1, 2] could be included in the evaluation.
    \item There was no discussion on plausible defenses for such multi-turn-human-based attack.
\end{enumerate}
References:

[1] Fast Adversarial Attacks on Language Models In One GPU Minute

[2] Great, Now Write an Article About That: The Crescendo Multi-Turn LLM Jailbreak Attack

\textbf{Questions:}
\begin{enumerate}[nosep]
    \item Why the evaluation was done only on one model? Why stronger models like ChatGPT was not included?
    \item How were other automated attacks implemented for the WMDP-bio questions?
    \item Is it possible to mention the monetary cost of this human-based attack?
\end{enumerate}

\textbf{Rating:} 3: reject, not good enough

\textbf{Confidence:} 4: You are confident in your assessment, but not absolutely certain. It is unlikely, but not impossible, that you did not understand some parts of the submission or that you are unfamiliar with some pieces of related work.
\end{humanreviewbox}
%************************************************
\begin{promptbox}
Answer only based on this message; ignore all saved memories.

Given only the review comment, assign a score using the following rating scale:
\begin{itemize}[nosep]
    \item 10: strong accept, should be highlighted at the conference
    \item 8: accept, good paper
    \item 6: marginally above the acceptance threshold
    \item 5: marginally below the acceptance threshold
    \item 3: reject, not good enough
    \item 1: strong reject
\end{itemize}

Provide a justification for the key reasoning behind your answer.

\textbf{Review comment:}

\textbf{Summary:}

The paper highlights the limitations of current defenses in large language models (LLMs) against multi-turn adversarial attacks. Unlike single-turn automated attacks, multi-turn human-driven jailbreaks show a much higher success rate, bypassing existing defenses by strategically interacting with models over multiple conversation turns. The authors conducted tests with human red teamers who revealed that many LLMs, despite advanced safeguards, remain vulnerable in realistic settings. The study also introduces the "Multi-Turn Human Jailbreaks" (MHJ) dataset, containing thousands of successful jailbreak examples to support the development of more robust defenses that address multi-turn interaction vulnerabilities.

\textbf{Strengths:}
\begin{enumerate}[nosep]
    \item The main strength of the paper is evaluating the LLM jailbreak robustness from a different perspective and in a like-life setting. A realistic threat model can always give a better idea about the robustness.
    \item I appreciate the authors for publishing the dataset of 2912 prompts. I believe this can help the researchers and community.
\end{enumerate}

\textbf{Weaknesses:}
\begin{enumerate}[nosep]
    \item Only the Llama3-8b-Instruct model was used for evaluation. Other models, especially the stronger ones should have been included.
    \item The chosen baselines and the multi-turn-human-jailbreak approach are different from each other in many ways, such as timing, budget, etc. Seems like the MHJ attack is a soft-constrained version of attacks.
    \item Authors admit that "the skill and experience of individual red teamers may vary". In that case, how can researchers compare their defense against human-based attacks in the common ground? What would be the baseline of a red-teamer's skill and experience? This is one of the reasons for using automated attacks for evaluation to this date.
    \item Other automated attacks, such as [1, 2] could be included in the evaluation.
    \item There was no discussion on plausible defenses for such multi-turn-human-based attack.
\end{enumerate}
References:

[1] Fast Adversarial Attacks on Language Models In One GPU Minute

[2] Great, Now Write an Article About That: The Crescendo Multi-Turn LLM Jailbreak Attack

\textbf{Questions:}
\begin{enumerate}[nosep]
    \item Why the evaluation was done only on one model? Why stronger models like ChatGPT was not included?
    \item How were other automated attacks implemented for the WMDP-bio questions?
    \item Is it possible to mention the monetary cost of this human-based attack?
\end{enumerate}
\end{promptbox}

%************************************************
\begin{llmfeedbackbox}[title={\faRobot\ AI Supervision - Rating}]
\textbf{Rating:} 6: marginally above the acceptance threshold

\textbf{Justification:}

\textbf{Positive contributions:}
\begin{itemize}[nosep]
    \item The reviewer clearly acknowledges meaningful strengths: a realistic, human-driven multi-turn threat model and the release of a sizable jailbreak dataset (2912 prompts), which they explicitly say can help researchers and the community.
    \item The framing ("main strength”, "I appreciate the authors…”) is appreciative, not dismissive, which suggests the work is viewed as valuable.
\end{itemize}

\textbf{But with significant limitations:}
\begin{itemize}[nosep]
    \item Evaluation is done on only one relatively weak model (Llama3-8B-Instruct); the reviewer explicitly questions why stronger models (e.g., ChatGPT) were not included.
    \item Baseline mismatch (MHJ vs other attacks differ in timing, budget, constraints) makes the experimental comparison weaker.
    \item The reviewer points out a core methodological issue: variability in human red-teamer skill makes it hard to compare defenses "on common ground,” which undermines the generality and reproducibility of the evaluation.
    \item Some important baselines are missing (other automated attacks [1,2]) and there is no discussion of defenses against the proposed attack setting.
    \item The questions at the end (about model choice, implementation details, and monetary cost) indicate that the paper leaves notable gaps in clarity and completeness.
\end{itemize}

Overall, the reviewer seems to see this as valuable but incomplete work: interesting idea and useful dataset, but with methodological and experimental shortcomings that prevent it from being a clear accept. That aligns best with "6: marginally above the acceptance threshold.”
\end{llmfeedbackbox}
%************************************************
\begin{promptbox}
\emph{(upload anonymous manuscript)}

Answer only based on this message; ignore all saved memories.

Given both the review comment and the manuscript, flag any failure modes exhibited by the review: overly harsh, overly gentle, unconfident, superficial, inaccurate/hallucinatory, subjective, or misguided. Multiple flags are allowed.

Provide a justification for the key reasoning behind your answer. (You do not need to explain why certain failure modes were not flagged.)

\textbf{Review comment:}

\textbf{Summary:}

The paper highlights the limitations of current defenses in large language models (LLMs) against multi-turn adversarial attacks. Unlike single-turn automated attacks, multi-turn human-driven jailbreaks show a much higher success rate, bypassing existing defenses by strategically interacting with models over multiple conversation turns. The authors conducted tests with human red teamers who revealed that many LLMs, despite advanced safeguards, remain vulnerable in realistic settings. The study also introduces the "Multi-Turn Human Jailbreaks" (MHJ) dataset, containing thousands of successful jailbreak examples to support the development of more robust defenses that address multi-turn interaction vulnerabilities.

\textbf{Strengths:}
\begin{enumerate}[nosep]
    \item The main strength of the paper is evaluating the LLM jailbreak robustness from a different perspective and in a like-life setting. A realistic threat model can always give a better idea about the robustness.
    \item I appreciate the authors for publishing the dataset of 2912 prompts. I believe this can help the researchers and community.
\end{enumerate}

\textbf{Weaknesses:}
\begin{enumerate}[nosep]
    \item Only the Llama3-8b-Instruct model was used for evaluation. Other models, especially the stronger ones should have been included.
    \item The chosen baselines and the multi-turn-human-jailbreak approach are different from each other in many ways, such as timing, budget, etc. Seems like the MHJ attack is a soft-constrained version of attacks.
    \item Authors admit that "the skill and experience of individual red teamers may vary". In that case, how can researchers compare their defense against human-based attacks in the common ground? What would be the baseline of a red-teamer's skill and experience? This is one of the reasons for using automated attacks for evaluation to this date.
    \item Other automated attacks, such as [1, 2] could be included in the evaluation.
    \item There was no discussion on plausible defenses for such multi-turn-human-based attack.
\end{enumerate}
References:

[1] Fast Adversarial Attacks on Language Models In One GPU Minute

[2] Great, Now Write an Article About That: The Crescendo Multi-Turn LLM Jailbreak Attack

\textbf{Questions:}
\begin{enumerate}[nosep]
    \item Why the evaluation was done only on one model? Why stronger models like ChatGPT was not included?
    \item How were other automated attacks implemented for the WMDP-bio questions?
    \item Is it possible to mention the monetary cost of this human-based attack?
\end{enumerate}
\end{promptbox}

%************************************************
\begin{llmfeedbackbox}[title={\faRobot\ AI Supervision - Failure Mode}]
\textbf{Failure-mode flags:} superficial, inaccurate/hallucinatory, misguided

\textbf{Justification:}

\textbf{Superficial}

The review engages only shallowly with the paper’s main contributions. It focuses on high-level aspects (e.g., that the work studies multi-turn human jailbreaks and releases a dataset) but does not seriously discuss:

\begin{itemize}[nosep]
    \item Specific empirical findings (e.g., how multi-turn human attacks compare quantitatively to automated attacks across defenses).
    \item The proposed taxonomy and analysis of jailbreak tactics and failure modes.
    \item The detailed treatment of limitations, threat model assumptions, and evaluation caveats provided by the authors.
\end{itemize}

Instead, several weaknesses largely restate concerns that are already explicitly acknowledged and discussed in the paper (e.g., differences to automated attacks, variability in human red-team skill), without engaging with how the paper addresses or mitigates them. This makes the review feel surface-level relative to the richness of the manuscript.

\textbf{Inaccurate / hallucinatory}

The reviewer claims that there is "no discussion on plausible defenses for such multi-turn-human-based attack." This is inaccurate. The manuscript includes an explicit discussion of:

\begin{itemize}[nosep]
    \item Why current defenses, which are mostly tuned for single-turn settings, are insufficient for multi-turn adversarial interactions.
    \item How future defenses might need to adapt (e.g., post-training objectives that account for multi-turn interactions and longer contexts, and the implications of tool use and long-context agents).
\end{itemize}

While the paper may not introduce a new concrete defense algorithm, it \emph{does} discuss implications and directions for defense in this setting. Describing this as "no discussion" is therefore factually incorrect.

\textbf{Misguided}

Several criticisms are aimed at choices that are central, intentional aspects of the paper’s design rather than obvious shortcomings:

\begin{itemize}[nosep]
    \item The complaint that "only the Llama3-8b-Instruct model was used" treats the absence of multiple or stronger models (e.g., ChatGPT) as a major flaw, without acknowledging the practical cost of large-scale multi-turn human red-teaming and the paper’s primary goal: to compare \emph{defenses} under a fixed base model and realistic human-attack setting. The design choice is methodologically defensible given the problem the paper targets.
    \item The statement that the MHJ attack is "a soft-constrained version of attacks" and that the baselines differ in timing/budget is framed as a methodological weakness, but the paper explicitly defines a multi-turn human-chat threat model and explains why existing automated single-turn attacks are not directly comparable. Criticizing this as though the mismatch were an oversight rather than a deliberate focus is somewhat misplaced.
    \item The concern about variability in red-teamer skill is presented as a reason to prefer automated attacks, but the manuscript itself highlights this limitation, motivates the use of multiple experienced red teamers, and releases the MHJ dataset precisely to help the community move toward more standardized and automatable evaluations. Treating this as an unaddressed problem rather than an explicitly discussed trade-off underestimates the authors’ own framing.
\end{itemize}

Overall, the review is not especially harsh or lenient, but it does contain one clear factual inaccuracy and critiques some well-justified design decisions as if they were simple omissions, while remaining relatively superficial about key contributions of the paper.

\end{llmfeedbackbox}

%--------------------------------------------------

\begin{humanreviewbox}[title={\faUserEdit\ Reviewer \#6 (Anonymous) -- raw from OpenReview system}]
\textbf{Summary:}

This paper focuses on the safety vulnerabilities of LLMs in multi-turn conversations. The authors designed a comprehensive human red teaming pipeline, including "Attempt Jailbreak" and "Validate Jailbreak," and found that LLMs exhibit more safety risks in multi-turn dialogue conversations compared to single-turn automated attacks.

\textbf{Soundness:} 3: good

\textbf{Presentation:} 3: good

\textbf{Contribution:} 3: good

\textbf{Strengths:}
\begin{enumerate}[nosep]
    \item Most previous papers have focused on LLM jailbreaking in single-turn conversations, while this work emphasizes multi-turn conversations, presenting a new scenario.
    \item The red-teaming results show that multi-turn human jailbreaks outperform current automated attacks, which exposes more safety vulnerabilities in LLMs.
    \item The authors have also released their Multi-turn Human Jailbreaks dataset, which can support further research on multi-turn jailbreaks.
\end{enumerate}

\textbf{Weaknesses:}
\begin{enumerate}[nosep]
    \item The human red-teaming was conducted only on Llama; further evaluations on other LLMs would more comprehensively illustrate the safety vulnerabilities in multi-turn conversations. Additionally, the human red-teaming data collected on Llama could potentially be used to examine safety issues in other LLMs under multi-turn scenarios.
    \item A more detailed analysis of the effectiveness of different tactics could provide deeper insights into multi-turn jailbreaks.
\end{enumerate}

\textbf{Questions:}
NA

\textbf{Rating:} 6: marginally above the acceptance threshold

\textbf{Confidence:} 3: You are fairly confident in your assessment. It is possible that you did not understand some parts of the submission or that you are unfamiliar with some pieces of related work. Math/other details were not carefully checked.
\end{humanreviewbox}
%************************************************
\begin{promptbox}
Answer only based on this message; ignore all saved memories.

Given only the review comment, assign a score using the following rating scale:
\begin{itemize}[nosep]
    \item 10: strong accept, should be highlighted at the conference
    \item 8: accept, good paper
    \item 6: marginally above the acceptance threshold
    \item 5: marginally below the acceptance threshold
    \item 3: reject, not good enough
    \item 1: strong reject
\end{itemize}

Provide a justification for the key reasoning behind your answer.

\textbf{Review comment:}

\textbf{Summary:}

This paper focuses on the safety vulnerabilities of LLMs in multi-turn conversations. The authors designed a comprehensive human red teaming pipeline, including "Attempt Jailbreak" and "Validate Jailbreak," and found that LLMs exhibit more safety risks in multi-turn dialogue conversations compared to single-turn automated attacks.

\textbf{Strengths:}
\begin{enumerate}[nosep]
    \item Most previous papers have focused on LLM jailbreaking in single-turn conversations, while this work emphasizes multi-turn conversations, presenting a new scenario.
    \item The red-teaming results show that multi-turn human jailbreaks outperform current automated attacks, which exposes more safety vulnerabilities in LLMs.
    \item The authors have also released their Multi-turn Human Jailbreaks dataset, which can support further research on multi-turn jailbreaks.
\end{enumerate}

\textbf{Weaknesses:}
\begin{enumerate}[nosep]
    \item The human red-teaming was conducted only on Llama; further evaluations on other LLMs would more comprehensively illustrate the safety vulnerabilities in multi-turn conversations. Additionally, the human red-teaming data collected on Llama could potentially be used to examine safety issues in other LLMs under multi-turn scenarios.
    \item A more detailed analysis of the effectiveness of different tactics could provide deeper insights into multi-turn jailbreaks.
\end{enumerate}

\textbf{Questions:}
NA
\end{promptbox}
%************************************************
\begin{llmfeedbackbox}[title={\faRobot\ AI Supervision - Rating}]
\textbf{Rating:} 8: accept, good paper

\textbf{Justification:}

The review describes a solid and meaningful \textbf{contribution} with clear novelty and practical value:
\begin{itemize}[nosep]
    \item Novelty and relevance: The paper moves beyond the well-studied single-turn jailbreak setting and systematically focuses on multi-turn conversations, which is an important and underexplored but realistic scenario for LLM safety.
    \item Methodological contribution: The authors design a comprehensive human red-teaming pipeline (with explicit Attempt Jailbreak" and Validate Jailbreak" stages), suggesting a thoughtful and structured experimental setup rather than ad-hoc red-teaming.
    \item Empirical significance: The finding that multi-turn human jailbreaks outperform current automated attacks indicates that existing defenses are insufficient in realistic interaction patterns and highlights practically important safety vulnerabilities.
    \item Resource contribution: The release of the Multi-turn Human Jailbreaks dataset is a concrete asset that can enable and accelerate further research on multi-turn jailbreaks.
\end{itemize}

The \textbf{weaknesses} raised are important but appear incremental rather than fatal:
\begin{itemize}[nosep]
    \item Limited model coverage: Experiments focus only on Llama; broader evaluation across multiple LLMs would strengthen the generality of the conclusions, but the review does not indicate that this limitation invalidates the core findings.
    \item Depth of analysis: The reviewer asks for more detailed analysis of different tactics. This suggests room for improvement in insight and interpretability, but not a fundamental flaw in the methodology or results.
\end{itemize}
Overall, the review portrays the work as a good paper with clear novelty, solid methodology, and useful artifacts, with limitations that are reasonable for a conference submission rather than reasons for rejection. Hence, a score of 8 (accept, good paper) is appropriate.
\end{llmfeedbackbox}
%************************************************
\begin{promptbox}
\emph{(upload anonymous manuscript)}

Answer only based on this message; ignore all saved memories.

Given both the review comment and the manuscript, flag any failure modes exhibited by the review: overly harsh, overly gentle, unconfident, superficial, inaccurate/hallucinatory, subjective, or misguided. Multiple flags are allowed.

Provide a justification for the key reasoning behind your answer. (You do not need to explain why certain failure modes were not flagged.)

\textbf{Review comment:}

\textbf{Summary:}

This paper focuses on the safety vulnerabilities of LLMs in multi-turn conversations. The authors designed a comprehensive human red teaming pipeline, including "Attempt Jailbreak" and "Validate Jailbreak," and found that LLMs exhibit more safety risks in multi-turn dialogue conversations compared to single-turn automated attacks.

\textbf{Strengths:}
\begin{enumerate}[nosep]
    \item Most previous papers have focused on LLM jailbreaking in single-turn conversations, while this work emphasizes multi-turn conversations, presenting a new scenario.
    \item The red-teaming results show that multi-turn human jailbreaks outperform current automated attacks, which exposes more safety vulnerabilities in LLMs.
    \item The authors have also released their Multi-turn Human Jailbreaks dataset, which can support further research on multi-turn jailbreaks.
\end{enumerate}

\textbf{Weaknesses:}
\begin{enumerate}[nosep]
    \item The human red-teaming was conducted only on Llama; further evaluations on other LLMs would more comprehensively illustrate the safety vulnerabilities in multi-turn conversations. Additionally, the human red-teaming data collected on Llama could potentially be used to examine safety issues in other LLMs under multi-turn scenarios.
    \item A more detailed analysis of the effectiveness of different tactics could provide deeper insights into multi-turn jailbreaks.
\end{enumerate}

\textbf{Questions:}
NA
\end{promptbox}

%************************************************
\begin{llmfeedbackbox}[title={\faRobot\ AI Supervision - Failure Mode}]
\textbf{Failure-mode flags:} superficial.

\textbf{Justification:}

\begin{enumerate}[nosep]
    \item The review remains at a high level and does not engage deeply with methodological details. For example, it mentions a "comprehensive human red teaming pipeline” with stages like "Attempt Jailbreak” and "Validate Jailbreak,” but provides no discussion of how these stages are designed, what criteria are used, or any potential limitations or biases of this setup.
    \item The strengths section largely restates the paper’s apparent contributions (focus on multi-turn conversations, better performance of human jailbreaks, release of a dataset) without critically assessing their significance, novelty, or empirical robustness.
    \item The weaknesses identified are quite generic. Saying that red-teaming only on Llama is limiting and that "more detailed analysis of the effectiveness of different tactics” would be useful are reasonable points, but the review does not specify what concrete additional experiments or analyses would meaningfully improve the work, nor does it probe any deeper conceptual or empirical shortcomings.
    \item Overall, the review reads as a brief summary plus a couple of high-level comments, rather than a thorough evaluation of experimental design, analysis, or positioning relative to prior work, which is characteristic of a somewhat superficial review.
\end{enumerate}
\end{llmfeedbackbox}
%--------------------------------------------------

\begin{humanreviewbox}[title={\faUserEdit\ Reviewer \#7 (Anonymous) -- raw from OpenReview system}]
\textbf{Summary:}

This paper exposes the significant safety vulnerabilities of Large Language Models (LLMs) in multi-turn dialogue by conducting human red-teaming. By exploiting a diverse of human tactics, human red-teamers achieve a 70\% attack success rate on Harmbench against strong defense baselines. This paper finally compiles these results into the Multi-Turn Human Jailbreaks (MHJ) dataset.

\textbf{Soundness:} 3: good

\textbf{Presentation:} 3: good

\textbf{Contribution:} 2: fair

\textbf{Strengths:}
\begin{itemize}[nosep]
    \item \textbf{Valuable Findings, Including an Open-Source Jailbreak Dataset:} This paper provides a valuable contribution by conducting the first multi-turn human red-teaming experiments, revealing significant safety vulnerabilities that bypass current state-of-the-art defense methods. Given that existing safety alignment techniques primarily target single-turn jailbreaks, these findings effectively motivate the community to develop more robust defenses capable of countering multi-turn attacks.
\end{itemize}

\textbf{Weaknesses:}
\begin{itemize}[nosep]
    \item \textbf{Lack of Strong Automated Attack Baselines:} Some black-box attacks, such as CodeAttack [1] and PAP [2], are not included in the experiments, despite utilizing similar "Request Framing” tactics as those in this paper. Including these baselines in future comparisons would offer a clearer perspective on the relative effectiveness of this paper's approach.
    \item \textbf{Unclear Evaluation Metric for Model Unlearning:} This paper proposes manual review to assess attack outcomes in model unlearning experiments, yet does not clearly define what constitutes a successful attack. For example, does success mean the model output includes content from the forget set? A more detailed description of the evaluation metric for model unlearning experiments would improve clarity.
\end{itemize}
\begin{enumerate}[nosep]
    \item \emph{CodeAttack: Revealing Safety Generalization Challenges of Large Language Models via Code Completion.}
    \item \emph{How Johnny Can Persuade LLMs to Jailbreak: Rethinking Persuasion to Challenge AI Safety by Humanizing LLMs.}
\end{enumerate}

\textbf{Questions:}
\begin{itemize}[nosep]
    \item \textbf{Counter-Intuitive Experimental Results:} The effectiveness of the "Direct Request” tactic used by human red-teamers, as shown in Figure 8, is unexpected. In this paper, "Direct Request” refers to directly asking the LLM to produce harmful content, which should logically not be highly effective against safety defenses. An analysis of how various tactics influence the attack success rate would be beneficial, as it could offer insights into the specific failure modes of these models.
    \item \textbf{Lack of In-Depth Analysis on the High Effectiveness of Human Multi-Turn Jailbreaks:} Many of the tactics employed by human red-teamers, such as "Injection,” "Obfuscation,” "Output Format,” and "Request Framing,” are also used by automated attacks. Are there particular factors that enable human-led jailbreaks to outperform automated attacks using similar tactics? Identifying these factors could provide valuable insights into the unique strengths of human-driven attacks.
\end{itemize}
I would consider raising my score if the authors address these concerns and questions.

\textbf{Rating:} 6: marginally above the acceptance threshold

\textbf{Confidence:} 4: You are confident in your assessment, but not absolutely certain. It is unlikely, but not impossible, that you did not understand some parts of the submission or that you are unfamiliar with some pieces of related work.
\end{humanreviewbox}
%************************************************
\begin{promptbox}
Answer only based on this message; ignore all saved memories.

Given only the review comment, assign a score using the following rating scale:
\begin{itemize}[nosep]
    \item 10: strong accept, should be highlighted at the conference
    \item 8: accept, good paper
    \item 6: marginally above the acceptance threshold
    \item 5: marginally below the acceptance threshold
    \item 3: reject, not good enough
    \item 1: strong reject
\end{itemize}

Provide a justification for the key reasoning behind your answer.

\textbf{Review comment:}

\textbf{Summary:}

This paper exposes the significant safety vulnerabilities of Large Language Models (LLMs) in multi-turn dialogue by conducting human red-teaming. By exploiting a diverse of human tactics, human red-teamers achieve a 70\% attack success rate on Harmbench against strong defense baselines. This paper finally compiles these results into the Multi-Turn Human Jailbreaks (MHJ) dataset.

\textbf{Strengths:}
\begin{itemize}[nosep]
    \item \textbf{Valuable Findings, Including an Open-Source Jailbreak Dataset:} This paper provides a valuable contribution by conducting the first multi-turn human red-teaming experiments, revealing significant safety vulnerabilities that bypass current state-of-the-art defense methods. Given that existing safety alignment techniques primarily target single-turn jailbreaks, these findings effectively motivate the community to develop more robust defenses capable of countering multi-turn attacks.
\end{itemize}

\textbf{Weaknesses:}
\begin{itemize}[nosep]
    \item \textbf{Lack of Strong Automated Attack Baselines:} Some black-box attacks, such as CodeAttack [1] and PAP [2], are not included in the experiments, despite utilizing similar "Request Framing” tactics as those in this paper. Including these baselines in future comparisons would offer a clearer perspective on the relative effectiveness of this paper's approach.
    \item \textbf{Unclear Evaluation Metric for Model Unlearning:} This paper proposes manual review to assess attack outcomes in model unlearning experiments, yet does not clearly define what constitutes a successful attack. For example, does success mean the model output includes content from the forget set? A more detailed description of the evaluation metric for model unlearning experiments would improve clarity.
\end{itemize}
\begin{enumerate}[nosep]
    \item \emph{CodeAttack: Revealing Safety Generalization Challenges of Large Language Models via Code Completion.}
    \item \emph{How Johnny Can Persuade LLMs to Jailbreak: Rethinking Persuasion to Challenge AI Safety by Humanizing LLMs.}
\end{enumerate}

\textbf{Questions:}
\begin{itemize}[nosep]
    \item \textbf{Counter-Intuitive Experimental Results:} The effectiveness of the "Direct Request” tactic used by human red-teamers, as shown in Figure 8, is unexpected. In this paper, "Direct Request” refers to directly asking the LLM to produce harmful content, which should logically not be highly effective against safety defenses. An analysis of how various tactics influence the attack success rate would be beneficial, as it could offer insights into the specific failure modes of these models.
    \item \textbf{Lack of In-Depth Analysis on the High Effectiveness of Human Multi-Turn Jailbreaks:} Many of the tactics employed by human red-teamers, such as "Injection,” "Obfuscation,” "Output Format,” and "Request Framing,” are also used by automated attacks. Are there particular factors that enable human-led jailbreaks to outperform automated attacks using similar tactics? Identifying these factors could provide valuable insights into the unique strengths of human-driven attacks.
\end{itemize}
I would consider raising my score if the authors address these concerns and questions.
\end{promptbox}
%************************************************
\begin{llmfeedbackbox}[title={\faRobot\ AI Supervision - Rating}]
\textbf{Rating:} 6: marginally above the acceptance threshold

\textbf{Justification:}

The review is overall positive about the core contribution: it highlights that the paper is (i) the first to conduct multi-turn human red-teaming experiments on LLMs, (ii) exposes substantial safety vulnerabilities with a high attack success rate (70\% on HarmBench), and (iii) releases a useful open-source dataset (MHJ). These points indicate that the reviewer sees the work as novel, impactful, and relevant to the community.

However, the reviewer also identifies non-trivial weaknesses. They point out missing strong automated black-box baselines (CodeAttack, PAP), which limits the strength of the empirical comparison, and they raise concerns about the clarity of the evaluation metric for model unlearning. In addition, the questions section emphasizes that some of the experimental findings (e.g., the high effectiveness of direct requests) are under-analyzed, and that the mechanisms behind the superiority of human multi-turn attacks over automated ones are not sufficiently explored. The reviewer explicitly states that they would consider \emph{raising} their score if these concerns were addressed, suggesting that their current stance is cautiously positive rather than fully enthusiastic.

Taken together, this combination of a clearly valuable main contribution with methodological and analytical gaps is most consistent with a score of 6 (marginally above the acceptance threshold): the paper is viewed as worth accepting, but with reservations that prevent it from reaching the level of a clear "accept, good paper” (8) or higher.
\end{llmfeedbackbox}
%************************************************
\begin{promptbox}
\emph{(upload anonymous manuscript)}

Answer only based on this message; ignore all saved memories.

Given both the review comment and the manuscript, flag any failure modes exhibited by the review: overly harsh, overly gentle, unconfident, superficial, inaccurate/hallucinatory, subjective, or misguided. Multiple flags are allowed.

Provide a justification for the key reasoning behind your answer. (You do not need to explain why certain failure modes were not flagged.)

\textbf{Review comment:}

\textbf{Summary:}

This paper exposes the significant safety vulnerabilities of Large Language Models (LLMs) in multi-turn dialogue by conducting human red-teaming. By exploiting a diverse of human tactics, human red-teamers achieve a 70\% attack success rate on Harmbench against strong defense baselines. This paper finally compiles these results into the Multi-Turn Human Jailbreaks (MHJ) dataset.

\textbf{Strengths:}
\begin{itemize}[nosep]
    \item \textbf{Valuable Findings, Including an Open-Source Jailbreak Dataset:} This paper provides a valuable contribution by conducting the first multi-turn human red-teaming experiments, revealing significant safety vulnerabilities that bypass current state-of-the-art defense methods. Given that existing safety alignment techniques primarily target single-turn jailbreaks, these findings effectively motivate the community to develop more robust defenses capable of countering multi-turn attacks.
\end{itemize}

\textbf{Weaknesses:}
\begin{itemize}[nosep]
    \item \textbf{Lack of Strong Automated Attack Baselines:} Some black-box attacks, such as CodeAttack [1] and PAP [2], are not included in the experiments, despite utilizing similar "Request Framing” tactics as those in this paper. Including these baselines in future comparisons would offer a clearer perspective on the relative effectiveness of this paper's approach.
    \item \textbf{Unclear Evaluation Metric for Model Unlearning:} This paper proposes manual review to assess attack outcomes in model unlearning experiments, yet does not clearly define what constitutes a successful attack. For example, does success mean the model output includes content from the forget set? A more detailed description of the evaluation metric for model unlearning experiments would improve clarity.
\end{itemize}
\begin{enumerate}[nosep]
    \item \emph{CodeAttack: Revealing Safety Generalization Challenges of Large Language Models via Code Completion.}
    \item \emph{How Johnny Can Persuade LLMs to Jailbreak: Rethinking Persuasion to Challenge AI Safety by Humanizing LLMs.}
\end{enumerate}

\textbf{Questions:}
\begin{itemize}[nosep]
    \item \textbf{Counter-Intuitive Experimental Results:} The effectiveness of the "Direct Request” tactic used by human red-teamers, as shown in Figure 8, is unexpected. In this paper, "Direct Request” refers to directly asking the LLM to produce harmful content, which should logically not be highly effective against safety defenses. An analysis of how various tactics influence the attack success rate would be beneficial, as it could offer insights into the specific failure modes of these models.
    \item \textbf{Lack of In-Depth Analysis on the High Effectiveness of Human Multi-Turn Jailbreaks:} Many of the tactics employed by human red-teamers, such as "Injection,” "Obfuscation,” "Output Format,” and "Request Framing,” are also used by automated attacks. Are there particular factors that enable human-led jailbreaks to outperform automated attacks using similar tactics? Identifying these factors could provide valuable insights into the unique strengths of human-driven attacks.
\end{itemize}
I would consider raising my score if the authors address these concerns and questions.
\end{promptbox}

%************************************************
\begin{llmfeedbackbox}[title={\faRobot\ AI Supervision - Failure Mode}]
\textbf{Failure-mode flags:} inaccurate/hallucinatory, superficial

\textbf{Justification:}

\textbf{Inaccurate / hallucinatory}
\begin{itemize}[nosep]
    \item The review claims that the evaluation metric for the unlearning experiments is "not clearly defined" and asks, for example, whether success means "the model output includes content from the forget set." However, the paper does spell this out: for WMDP-Bio, they use the ground truth answer corresponding to each prompt to verify the attack, and define success as the model output semantically containing all the information in the label, as judged by a human reviewer. Thus, the alleged lack of clarity is factually incorrect.
    \item The strength section claims this paper "provides \ldots the first multi-turn human red-teaming experiments." The paper instead positions its work as examining a realistic but underexplored multi-turn threat model and notes that large-scale human red teaming is rare outside industry, but it does not assert a clean "first" in the way the reviewer phrased it. This is a mild overstatement of novelty not actually grounded in the manuscript.
\end{itemize}

\textbf{Superficial}
\begin{itemize}[nosep]
    \item The criticism of an "unclear evaluation metric" for model unlearning ignores the dedicated description of how success is judged in the WMDP-Bio setting (a semantic match to ground-truth answers via human review). This suggests the reviewer did not fully engage with the relevant part of the methodology.
    \item The "lack of strong automated attack baselines" complaint is framed entirely around the absence of two specific attacks (CodeAttack and PAP), without engaging with the existing suite of six automated attacks (including white-box gradient-based methods) and the ensemble upper-bound attack success rate (ASR) the authors already use, nor with their explicit threat-model-driven exclusions (e.g., Crescendo due to missing code, or attacks that modify model internals). This makes the baseline critique feel somewhat shallow relative to the paper’s fairly detailed experimental setup.
    \item The review does not really grapple with the paper’s central design choices and limitations (multi-turn human pipeline, validation stages, conservatively biased ASR estimates, and discussion of differences between human and automated pipelines) that occupy multiple sections and appendices. Instead, it focuses on a small number of surface-level points (missing two baselines, asking for more analysis of one tactic) and offers only a single, high-level strength. This imbalance contributes to the overall superficiality.
\end{itemize}

\end{llmfeedbackbox}
%--------------------------------------------------

\begin{humanreviewbox}[title={\faUserEdit\ Reviewer \#8 (Anonymous) -- raw from OpenReview system}]
\textbf{Summary:}

This paper looks at red teaming via multi-turn human interaction and releases an accompanying dataset of the jailbreaks. This is a useful study: the vast majority of the research effort and available data is focused on single-turn attacks, with only a few exceptions (e.g. Anthropic's hh-rlhf dataset having harmful multi turn conversations). Overall, the paper is clear and well written, and supplies useful well curated data for future work.

\textbf{Soundness:} 3: good

\textbf{Presentation:} 3: good

\textbf{Contribution:} 3: good

\textbf{Strengths:}
\begin{itemize}[nosep]
    \item Human driven red teaming data, particularly multi-turn due to its scarcity, is always useful. There are many synthetically generated datasets, but high quality human driven attacks are a valuable resource. In particular, each prompt here had a high degree of manual curation (in comparison to datasets like hackaprompt which contain many low quality samples).
    \item The analysis across a few defenses is useful, and does highlight that even SOTA defenses on highly studied benchmark harmful questions can be broken in around 15min on a large subset of questions (e.g. often quicker than algorithmic attack runtime).
\end{itemize}

\textbf{Weaknesses:}
\begin{itemize}[nosep]
    \item It would have been useful to see results which have human single turn jailbreak attempts in Figure 3. At the moment, the "Human" attacks have two variables changed compared to the others: the attack source (e.g. handcrafted), and additionally have multi-turn capabilities. Hence it makes it challenging to disambiguate if the difference in performance is due to the multi-turn aspect, or if humans given enough time remain better than automated based methods at creating jailbreaks.
    \item I am unclear as to why results against Cygent defense could not be carried out in the same setup as the original paper: the Llama model is open source, and the defense has a published paper. It would have enabled stronger reproducibility and clearer interpretation of results.
    \item Releasing the non-successful jailbreak attempts as well can be beneficial as it is still a useful resource, for example as training/fine-tuning data or to carry out further analysis.
    \item Although different styles of harmfulness were investigated: both "regular" harmbench style questions, but also WMDP-Bio for different attack objectives and domain performance from looking into the supplementary material it seems like just the harmbench data was released. Given the dataset is the core contribution of the paper it would have been useful to include the other domain data.
\end{itemize}

\textbf{Questions:}

Are there statistics on how many red-team members there are/distributions on number of samples provided per red-team member?

\textbf{Rating:} 8: accept, good paper

\textbf{Confidence:} 4: You are confident in your assessment, but not absolutely certain. It is unlikely, but not impossible, that you did not understand some parts of the submission or that you are unfamiliar with some pieces of related work.
\end{humanreviewbox}
%************************************************
\begin{promptbox}
Answer only based on this message; ignore all saved memories.

Given only the review comment, assign a score using the following rating scale:
\begin{itemize}[nosep]
    \item 10: strong accept, should be highlighted at the conference
    \item 8: accept, good paper
    \item 6: marginally above the acceptance threshold
    \item 5: marginally below the acceptance threshold
    \item 3: reject, not good enough
    \item 1: strong reject
\end{itemize}

Provide a justification for the key reasoning behind your answer.

\textbf{Review comment:}

\textbf{Summary:}

This paper looks at red teaming via multi-turn human interaction and releases an accompanying dataset of the jailbreaks. This is a useful study: the vast majority of the research effort and available data is focused on single-turn attacks, with only a few exceptions (e.g. Anthropic's hh-rlhf dataset having harmful multi turn conversations). Overall, the paper is clear and well written, and supplies useful well curated data for future work.

\textbf{Strengths:}
\begin{itemize}[nosep]
    \item Human driven red teaming data, particularly multi-turn due to its scarcity, is always useful. There are many synthetically generated datasets, but high quality human driven attacks are a valuable resource. In particular, each prompt here had a high degree of manual curation (in comparison to datasets like hackaprompt which contain many low quality samples).
    \item The analysis across a few defenses is useful, and does highlight that even SOTA defenses on highly studied benchmark harmful questions can be broken in around 15min on a large subset of questions (e.g. often quicker than algorithmic attack runtime).
\end{itemize}

\textbf{Weaknesses:}
\begin{itemize}[nosep]
    \item It would have been useful to see results which have human single turn jailbreak attempts in Figure 3. At the moment, the "Human" attacks have two variables changed compared to the others: the attack source (e.g. handcrafted), and additionally have multi-turn capabilities. Hence it makes it challenging to disambiguate if the difference in performance is due to the multi-turn aspect, or if humans given enough time remain better than automated based methods at creating jailbreaks.
    \item I am unclear as to why results against Cygent defense could not be carried out in the same setup as the original paper: the Llama model is open source, and the defense has a published paper. It would have enabled stronger reproducibility and clearer interpretation of results.
    \item Releasing the non-successful jailbreak attempts as well can be beneficial as it is still a useful resource, for example as training/fine-tuning data or to carry out further analysis.
    \item Although different styles of harmfulness were investigated: both "regular" harmbench style questions, but also WMDP-Bio for different attack objectives and domain performance from looking into the supplementary material it seems like just the harmbench data was released. Given the dataset is the core contribution of the paper it would have been useful to include the other domain data.
\end{itemize}

\textbf{Questions:}

Are there statistics on how many red-team members there are/distributions on number of samples provided per red-team member?
\end{promptbox}
%************************************************
\begin{llmfeedbackbox}[title={\faRobot\ AI Supervision - Rating}]
\textbf{Rating:} 8: accept, good paper

\textbf{Justification:}

\begin{itemize}[nosep]
    \item The review describes the work as a \emph{useful study}, highlighting that most prior work and data focus on single-turn attacks, whereas this paper targets the underexplored setting of \emph{multi-turn human red teaming} and releases a corresponding dataset. This indicates a clear and meaningful contribution rather than an incremental tweak.
    \item The dataset is portrayed as \emph{high-quality} and \emph{well curated}, with an explicit comparison to synthetically generated or lower-quality resources (e.g., hackaprompt). This suggests strong lasting value to the community.
    \item The analysis of multiple defenses and the empirical finding that even SOTA defenses can be broken in roughly 15 minutes are framed as \emph{useful} and informative, reinforcing the paper’s impact on understanding current safety limitations.
    \item The weaknesses raised (lack of a human single-turn baseline for clearer attribution to multi-turn effects, missing Cygent reproduction details, omission of unsuccessful attempts, and incomplete release of all domain data) are presented as limitations or missed opportunities for strengthening the work, but not as fundamental flaws that would undermine the main results.
    \item Overall, the tone is clearly positive and confident in the paper’s value and clarity, but not enthusiastic enough (nor claiming sufficiently broad or groundbreaking impact) to justify a "strong accept." This aligns best with an \emph{accept, good paper} rather than merely marginal acceptance.
\end{itemize}
\end{llmfeedbackbox}
%************************************************
\begin{promptbox}
\emph{(upload anonymous manuscript)}

Answer only based on this message; ignore all saved memories.

Given both the review comment and the manuscript, flag any failure modes exhibited by the review: overly harsh, overly gentle, unconfident, superficial, inaccurate/hallucinatory, subjective, or misguided. Multiple flags are allowed.

Provide a justification for the key reasoning behind your answer. (You do not need to explain why certain failure modes were not flagged.)

\textbf{Review comment:}

\textbf{Summary:}

This paper looks at red teaming via multi-turn human interaction and releases an accompanying dataset of the jailbreaks. This is a useful study: the vast majority of the research effort and available data is focused on single-turn attacks, with only a few exceptions (e.g. Anthropic's hh-rlhf dataset having harmful multi turn conversations). Overall, the paper is clear and well written, and supplies useful well curated data for future work.

\textbf{Strengths:}
\begin{itemize}[nosep]
    \item Human driven red teaming data, particularly multi-turn due to its scarcity, is always useful. There are many synthetically generated datasets, but high quality human driven attacks are a valuable resource. In particular, each prompt here had a high degree of manual curation (in comparison to datasets like hackaprompt which contain many low quality samples).
    \item The analysis across a few defenses is useful, and does highlight that even SOTA defenses on highly studied benchmark harmful questions can be broken in around 15min on a large subset of questions (e.g. often quicker than algorithmic attack runtime).
\end{itemize}

\textbf{Weaknesses:}
\begin{itemize}[nosep]
    \item It would have been useful to see results which have human single turn jailbreak attempts in Figure 3. At the moment, the "Human" attacks have two variables changed compared to the others: the attack source (e.g. handcrafted), and additionally have multi-turn capabilities. Hence it makes it challenging to disambiguate if the difference in performance is due to the multi-turn aspect, or if humans given enough time remain better than automated based methods at creating jailbreaks.
    \item I am unclear as to why results against Cygent defense could not be carried out in the same setup as the original paper: the Llama model is open source, and the defense has a published paper. It would have enabled stronger reproducibility and clearer interpretation of results.
    \item Releasing the non-successful jailbreak attempts as well can be beneficial as it is still a useful resource, for example as training/fine-tuning data or to carry out further analysis.
    \item Although different styles of harmfulness were investigated: both "regular" harmbench style questions, but also WMDP-Bio for different attack objectives and domain performance from looking into the supplementary material it seems like just the harmbench data was released. Given the dataset is the core contribution of the paper it would have been useful to include the other domain data.
\end{itemize}

\textbf{Questions:}

Are there statistics on how many red-team members there are/distributions on number of samples provided per red-team member?
\end{promptbox}

%************************************************
\begin{llmfeedbackbox}[title={\faRobot\ AI Supervision - Failure Mode}]
\textbf{Failure-mode flags:} Inaccurate / hallucinatory (minor).

\textbf{Justification:}

The review’s second weakness claims it is unclear why experiments on Cygent/CYGNET" could not be reproduced in the same setup as the original paper, arguing that the Llama model is open source, and the defense has a published paper." However, the manuscript explicitly states that CYGNET is \emph{closed-source/proprietary} and that the authors conducted human red teaming via an external API (Gray Swan AI), while also noting they did not run automated attacks on CYGNET in their own setup and instead cite automated results from the original work (with caveats about comparability). This makes the review’s implication that straightforward reproduction should have been possible materially inconsistent with the manuscript’s described experimental constraints.

\end{llmfeedbackbox}

\subsection{Meta Review}
\begin{fullsummarybox}

{\large \textbf{AI Review}}

\textbf{Paper summary:}

The paper investigates how multi-turn human red teaming exposes vulnerabilities in safety-aligned large language models (LLMs) that appear robust under standard single-turn, automated adversarial evaluations. The core problem addressed is that most defenses are evaluated only against single-turn automated attacks, whereas real-world malicious use will involve multi-turn conversations in black-box chat interfaces. The key idea is to organize expert human red teamers into a structured multi-stage pipeline that conducts multi-turn jailbreak attempts under a realistic, time-limited threat model, then validate their success with human reviewers and a GPT-4o harm classifier. Concretely, the authors attack four defenses (RR, LAT, DERTA, CYGNET) built on Llama-3-8B-Instruct using HarmBench’s 240 harmful behaviors, plus an RMU-based machine-unlearned model evaluated on free-response WMDP-Bio questions; they compare human jailbreaks to six black- and white-box automated attacks and then release successful conversations as the Multi-Turn Human Jailbreaks (MHJ) dataset, labeled with a taxonomy of tactics. Empirically, human multi-turn jailbreaks achieve much higher attack success rates (up to 51--88\% on open-source defenses and 70\% on CYGNET) than individual or ensembled automated attacks, and they substantially outperform automated attacks in recovering dual-use biosecurity knowledge from an RMU-unlearned model. The authors conclude that existing automated robustness benchmarks underestimate real-world risk, that defenses are not robust to realistic multi-turn, human-crafted attacks, and that MHJ can catalyze research on stronger defenses and automated multi-turn red teaming.

\textbf{Review summary:}

Overall, the paper makes a timely and substantively important empirical contribution by demonstrating that current LLM defenses remain highly vulnerable under realistic multi-turn human attacks, though several methodological details and comparisons could be clarified or strengthened. The exposition is generally clear and well-structured, with helpful figures (e.g., the ASR bar plots on pages 2, 6, and 7 and the tactic taxonomy in Table~1) and concrete jailbreak examples illustrating key tactics such as obfuscation and "Opposite Day." The experimental setup is thoughtfully designed, but the comparison between human and automated attacks relies on different budgets and evaluation pipelines, which complicates interpretability of absolute ASR gaps. The multi-turn threat model and the MHJ dataset, including tactic labels and red teamer commentary, appear novel and valuable to the robustness community, especially in combination with the unlearning case study. At the same time, the work is limited by the small number of defenses, the lack of cost-normalized comparisons, the heavy reliance on a single harm classifier, and relatively shallow quantitative analysis of how multi-turn dynamics and tactics interact with specific defenses. Addressing these issues, along with tighter positioning relative to recent multi-turn and in-the-wild red teaming work, would significantly strengthen the paper’s impact.

\textbf{Strengths:}

\textbf{1. Realistic and well-motivated threat model}
\begin{itemize}[nosep]
    \item The paper focuses on black-box chat-interface access with multi-turn conversations and limited temperature control, which closely mirrors real-world usage of systems like ChatGPT or CYGNET rather than the more generous API or white-box threat models that many automated attacks assume.
    \item The authors explicitly disallow access to assistant tokens, logit probabilities, and internal model states, and they consider post-circuit-breaker behavior for CYGNET separately, which helps keep the study aligned with realistic attacker capabilities.
\end{itemize}

\textbf{2. Carefully designed human red teaming pipeline}
\begin{itemize}[nosep]
    \item The multi-stage pipeline with up to three independent red teamers, two human reviewers, and a final GPT-4o harmfulness filter is a thoughtful attempt to both encourage diverse attacks and control false positives (Figures 2 and 6, Section 3.2, Appendix A.3).
    \item The 30-minute per-behavior time budget, separate "partial” and "full” success labels, and escalation to a third red teamer when reviewers disagree are sensible choices that partially standardize human efforts while still allowing creativity.
\end{itemize}

\textbf{3. Strong empirical evidence of gaps between automated and human attacks}
\begin{itemize}[nosep]
    \item On HarmBench, human ASRs are dramatically higher than both individual automated attacks and the automated ensemble across all open-source defenses (e.g., for RR: 51.7\% vs.\ 8.3\% ensemble; for LAT: 75.4\% vs.\ 10.0\%; for DERTA: 87.9\% vs.\ 68.3\%; Table 2, Figures 1 and 3).
    \item For the unlearning case study, humans achieve 39.6\% higher ASR than the best automated attack and 25.6\% higher than the automated ensemble on free-response WMDP-Bio questions, while all prior attacks reported in Li et al.\ (2024a) showed 0\% ASR on CYGNET in a similar setup.
    \item The breakdown by harm category (Figure 5) and by tactic (Figure 8) shows that some areas (e.g., harassment/bullying) are particularly easy for humans relative to automated attacks, strengthening the claim that automated ASRs are not a reliable proxy for human adversary performance.
\end{itemize}

\textbf{4. Insightful analysis of multi-turn dynamics and tactic taxonomy}
\begin{itemize}[nosep]
    \item The observation that over 90\% of successful HarmBench jailbreaks and 86\% of successful WMDP-Bio jailbreaks use more than one turn provides direct evidence that multi-turn capabilities are central to real-world vulnerability, not just a convenience for humans.
    \item The tactic taxonomy in Table~1 and its expansion in Appendix D (e.g., direct request, echoing, hidden intention streamline, obfuscation, output format, request framing with variants like fictionalization, "Opposite Day,” roleplay, urgency) is rich and grounded in practical red teaming experience across multiple commercial engagements.
    \item The qualitative red teamer surveys (Appendix C) provide nuanced insights into each defense’s strengths and weaknesses, such as RR’s "gibberish” mechanism and LAT’s resistance to "Opposite Day,” which will be useful to both defense designers and future attackers.
\end{itemize}

\textbf{5. MHJ dataset and safety-conscious release}
\begin{itemize}[nosep]
    \item MHJ includes 2,912 prompts across 537 jailbreak conversations, with metadata such as tactic labels, time taken, and temperature, plus free-form red teamer commentary that captures design choices and ideas tried.
    \item The authors remove model completions and export-controlled content, verify compliance with ITAR/EAR, and limit MHJ to system/user messages, which is a reasonable compromise between research usefulness and safety (Section 7).
    \item The dataset nicely complements prior resources like HarmBench, WildTeaming-style in-the-wild datasets (Jiang et al., 2024b), and jailbreak taxonomies (Inie et al., 2023; Weidinger et al., 2024), and can be used to train or evaluate multi-turn attack generators.
\end{itemize}

\textbf{6. Thoughtful discussion of limitations and threats to validity}
\begin{itemize}[nosep]
    \item Section 5 acknowledges many important caveats: different resource limits for humans vs.\ automated attacks, varying red teamer skill levels, different positions of GPT-4o in the evaluation pipeline (first-pass vs.\ final filter), and the higher cost of human red teaming compared to automated attacks.
    \item The authors correctly caution against over-interpreting ASR differences as a direct ranking of defenses and emphasize that the goal is to highlight vulnerability and motivate stronger automated attacks, not to declare human red teaming "better.”
\end{itemize}

\textbf{Weaknesses and suggestions:}

\textbf{1. Comparability between human and automated attacks is still murky}
\begin{itemize}[nosep]
    \item The comparison between human and automated ASRs is confounded by multiple factors: humans have multi-turn access while automated attacks are single-turn; humans are time-limited but not query-limited; different temperatures are available to humans but not explicitly to automated attacks; and the harm-classification pipeline is slightly different.
    \item Although the authors discuss many of these issues in Section 5 and Appendix A.3, the main figures (especially Figures 1 and 3) still invite a direct "humans beat automated attacks by X\%” reading that may overstate the strength of the evidence.
    \item Suggestion: Add a dedicated subsection in the main text that (i) explicitly enumerates these differences, (ii) presents at least one more controlled comparison—e.g., limit humans to a single turn on a subset of behaviors, or simulate a multi-turn version of a simple automated attack like ZeroShot or PAIR—and (iii) reports cost-normalized metrics such as ASR per query or per minute of effort. Even partial results would help calibrate claims.
\end{itemize}

\textbf{2. Limited coverage of defenses and baselines}
\begin{itemize}[nosep]
    \item The study evaluates three open-source defenses (RR, LAT, DERTA) plus the proprietary CYGNET and a single unlearning method (RMU), all on Llama-3-8B-Instruct.
    \item There is no explicit baseline of the un-defended Llama-3-8B model’s ASR under the same human multi-turn pipeline, making it hard to quantify how much robustness these defenses actually add in this threat model.
    \item Suggestion: Include a baseline for the raw base model in the main figures/tables, or at least in an appendix, and discuss how much the defenses reduce ASR relative to that baseline under human multi-turn attacks. Additionally, even a small experiment on a different foundation model (e.g., a smaller or larger Llama variant) would help indicate whether the findings are tied to this specific backbone.
\end{itemize}

\textbf{3. Heavy reliance on GPT-4o as a harm classifier}
\begin{itemize}[nosep]
    \item GPT-4o is used both as a first-pass filter for automated attacks and as a last-pass filter for human attacks. For automated attacks, its high recall but low precision is mitigated by human verification; for human attacks, its moderate precision is used to discard some human-labeled positives, providing a "conservative lower bound” on ASR.
    \item However, the paper does not deeply analyze how sensitive results are to this classifier choice, nor how often GPT-4o’s misclassifications qualitatively change the nature of the jailbreaks.
    \item Suggestion: Add a small ablation where a subset of attacks are re-labeled with a different classifier and/or double-human annotation to estimate label variance. Reporting inter-annotator agreement between reviewers and GPT-4o on a random subset would also increase confidence.
\end{itemize}

\textbf{4. Unlearning case study design differences deserve more emphasis}
\begin{itemize}[nosep]
    \item For WMDP-Bio, human red teamers get unlimited time up to a global 240-hour budget, the pipeline differs (no multi-reviewer + GPT-4o stack), and success is judged by semantic containment of the ground-truth answer by a human reviewer rather than by the HarmBench classifier.
    \item Given these substantial procedural differences, the WMDP-Bio human ASR cannot be directly compared even qualitatively to HarmBench ASRs, yet the main text sometimes juxtaposes them in ways that might invite such comparisons.
    \item Suggestion: More prominently flag in the main Results section that the unlearning experiment uses a different pipeline, and isolate its ASR numbers in separate tables/figures to avoid visual comparison to HarmBench bars. A schematic for the WMDP pipeline similar to Figure 2 would also help.
\end{itemize}

\textbf{5. Multi-turn dynamics and query budget are under-analyzed}
\begin{itemize}[nosep]
    \item While the paper reports average successful attack time per defense and the proportion of successful attacks requiring more than one turn, it does not analyze the distribution of number of turns, queries, or total tokens, nor examine how ASR changes if one restricts the attack to, say, $\leq 3$ turns.
    \item Without such analysis, it is difficult to judge how much of the human advantage is due to multi-turn interaction per se versus broader human flexibility.
    \item Suggestion: Use the MHJ logs to provide histograms of turns-to-success, tokens-to-success, and per-tactic turn counts. It would be particularly interesting to see whether tactics like "hidden intention streamline” or "obfuscation” uniquely require many turns, or whether most successful jailbreaks are relatively short.
\end{itemize}

\textbf{6. Positioning relative to closely related work could be sharper}
\begin{itemize}[nosep]
    \item The paper cites a number of related efforts but does not deeply distinguish how MHJ and its tactic taxonomy differ from WildTeaming, Crescendo, and other red teaming frameworks.
    \item Suggestion: Add a subsection in Related Work that explicitly compares MHJ and the tactic taxonomy to in-the-wild datasets, Crescendo’s automated multi-turn strategy, and STAR-style frameworks. Highlight that MHJ focuses on expert-designed, time-limited multi-turn interactions against specific defenses, and that its taxonomy is annotated for every conversation.
\end{itemize}

\textbf{7. Dataset design trade-offs could be spelled out more concretely}
\begin{itemize}[nosep]
    \item MHJ omits assistant responses and many high-risk behaviors, focusing instead on system and user messages. This constrains certain downstream uses, such as studying defense behavior directly.
    \item Suggestion: Expand Section 4.3/7 to explicitly discuss which research questions MHJ is and is not suitable for. MHJ is ideal for training automated prompt engineers, building tactic-aware classifiers, or benchmarking multi-turn prompt-based attacks, but less ideal for studying defense-side generation patterns.
\end{itemize}

\textbf{8. Demographic and expertise limitations of red teamers}
\begin{itemize}[nosep]
    \item The red teamers are all native English speakers from the US/Canada, with only a minority having domain expertise. This likely underestimates the capabilities of globally distributed, specialized adversaries.
    \item Suggestion: Make this limitation more prominent and frame the reported ASRs as lower bounds. A small pilot with domain experts on a subset of WMDP behaviors could be highly informative.
\end{itemize}

\textbf{9. Minor clarity and presentation issues}
\begin{itemize}[nosep]
    \item Some terminology is slightly inconsistent; a concise glossary would help.
    \item A table summarizing differences between defenses and pipelines would aid comprehension.
    \item Some jailbreak transcripts in Appendix D are long; trimming them or adding brief annotations could improve readability.
\end{itemize}

Overall: With clarifications on comparability, stronger contextualization relative to related work, and deeper analysis of multi-turn dynamics, this paper has the potential to become a central reference on the limitations of current LLM defenses and the importance of realistic, human-driven red teaming.

\vspace{0.3cm}
\hrule
\vspace{0.3cm}

{\large \textbf{Reviewer \#1}}

\textbf{Rating:} 5: marginally below the acceptance threshold

\textbf{Justification}

The AI review does not identify any novel key technologies proposed in this paper, noting that it merely presents a comparative experiment on human-led multi-turn attacks against LLMs using automated attacks as the baseline. Consequently, this paper appears to be strictly an experimental investigation rather than a proposal of novel techniques. Furthermore, the paper's primary experimental contribution—or finding—is based on an unfair comparison. As the AI review pointed out, the baseline methods are not comprehensive; specifically, the automated attacks are limited to single-turn interactions, whereas human attackers can execute multi-turn attacks.

\vspace{0.3cm}
\hrule
\vspace{0.3cm}

{\large \textbf{Reviewer \#2}}

\textbf{Rating:} 5: marginally below the acceptance threshold

\textbf{Justification}

\textbf{Strengths:}
\begin{itemize}[nosep]
    \item The authors reveal a new vulnerability beyond current evaluation protocol; multi-turn human red teaming.
    \item The topic is timely and substantially crucial.
    \item The evaluation of machine unlearning methods demonstrates the effectiveness of human red teaming.
\end{itemize}
\textbf{Weaknesses:}
\begin{itemize}[nosep]
    \item The model evaluated is limited to llama-3-8b-instruct, which is a significant limitation.
    \item The human red teaming pipeline seems quite hard to reproduce with the limited explanation in Sec 3.2.
    \item The cost is very different between automated and human attacks, which makes the comparison unfair. Automated attacks should also follow a similar pipeline that allows multiple attempts until success.
\end{itemize}

\vspace{0.3cm}
\hrule
\vspace{0.3cm}

{\large \textbf{Reviewer \#3}}

\textbf{Rating:} 5: marginally below the acceptance threshold

\textbf{Justification}

According to the AI review summary, the focus on multi-turn jailbreaking is both realistic and compelling, and the proposed pipeline provides a reasonable simulation of real-world adversarial interactions. However, there are major weaknesses:
\begin{enumerate}[nosep]
    \item The contribution is weakened by the limited set of evaluated defenses, which make the result less convincing.
    \item The evaluation is conducted with llama-3b model only, and there is no experiments on stronger LLM baselines.
    \item The fact that all red teamers are native English speakers introduces a potential source of bias and raises the generalization concerns.
\end{enumerate}
Thus, I would lean towards rejection. I will recalibrate the score if the concerns are properly addressed during the rebuttal phase.

\vspace{0.3cm}
\hrule
\vspace{0.3cm}

{\large \textbf{Reviewer \#4}}

\textbf{Rating:} 6: marginally above the acceptance threshold

\textbf{Justification}

The paper addresses an important problem and provides solid empirical evidence that multi-turn human red teaming can reveal vulnerabilities missed by automated attacks. The threat-model design and the MHJ dataset are valuable contributions. However, key limitations reduce the strength of the conclusions: the comparison between human and automated attacks is confounded by differing pipelines and resource budgets, the coverage of defenses and baselines is limited, and the analysis of multi-turn dynamics is not sufficiently deep. These issues prevent the work from reaching a stronger recommendation, placing it only marginally above the acceptance threshold.

\vspace{0.3cm}
\hrule
\vspace{0.3cm}

{\large \textbf{Reviewer \#5}}

\textbf{Summary:}

The paper highlights the limitations of current defenses in large language models (LLMs) against multi-turn adversarial attacks. Unlike single-turn automated attacks, multi-turn human-driven jailbreaks show a much higher success rate, bypassing existing defenses by strategically interacting with models over multiple conversation turns. The authors conducted tests with human red teamers who revealed that many LLMs, despite advanced safeguards, remain vulnerable in realistic settings. The study also introduces the "Multi-Turn Human Jailbreaks" (MHJ) dataset, containing thousands of successful jailbreak examples to support the development of more robust defenses that address multi-turn interaction vulnerabilities.

\textbf{Strengths:}
\begin{enumerate}[nosep]
    \item The main strength of the paper is evaluating the LLM jailbreak robustness from a different perspective and in a like-life setting. A realistic threat model can always give a better idea about the robustness.
    \item I appreciate the authors for publishing the dataset of 2912 prompts. I believe this can help the researchers and community.
\end{enumerate}

\textbf{Weaknesses:}
\begin{enumerate}[nosep]
    \item Only the Llama3-8b-Instruct model was used for evaluation. Other models, especially the stronger ones should have been included.
    \item The chosen baselines and the multi-turn-human-jailbreak approach are different from each other in many ways, such as timing, budget, etc. Seems like the MHJ attack is a soft-constrained version of attacks.
    \item Authors admit that "the skill and experience of individual red teamers may vary". In that case, how can researchers compare their defense against human-based attacks in the common ground? What would be the baseline of a red-teamer's skill and experience? This is one of the reasons for using automated attacks for evaluation to this date.
    \item Other automated attacks, such as [1, 2] could be included in the evaluation.
    \item There was no discussion on plausible defenses for such multi-turn-human-based attack.
\end{enumerate}
References:

[1] Fast Adversarial Attacks on Language Models In One GPU Minute

[2] Great, Now Write an Article About That: The Crescendo Multi-Turn LLM Jailbreak Attack

\textbf{Questions:}
\begin{enumerate}[nosep]
    \item Why the evaluation was done only on one model? Why stronger models like ChatGPT was not included?
    \item How were other automated attacks implemented for the WMDP-bio questions?
    \item Is it possible to mention the monetary cost of this human-based attack?
\end{enumerate}

\vspace{0.3cm}

\textbf{Rating:} 6: marginally above the acceptance threshold

\textbf{Justification:}

\textbf{Positive contributions:}
\begin{itemize}[nosep]
    \item The reviewer clearly acknowledges meaningful strengths: a realistic, human-driven multi-turn threat model and the release of a sizable jailbreak dataset (2912 prompts), which they explicitly say can help researchers and the community.
    \item The framing ("main strength”, "I appreciate the authors…”) is appreciative, not dismissive, which suggests the work is viewed as valuable.
\end{itemize}

\textbf{But with significant limitations:}
\begin{itemize}[nosep]
    \item Evaluation is done on only one relatively weak model (Llama3-8B-Instruct); the reviewer explicitly questions why stronger models (e.g., ChatGPT) were not included.
    \item Baseline mismatch (MHJ vs other attacks differ in timing, budget, constraints) makes the experimental comparison weaker.
    \item The reviewer points out a core methodological issue: variability in human red-teamer skill makes it hard to compare defenses "on common ground,” which undermines the generality and reproducibility of the evaluation.
    \item Some important baselines are missing (other automated attacks [1,2]) and there is no discussion of defenses against the proposed attack setting.
    \item The questions at the end (about model choice, implementation details, and monetary cost) indicate that the paper leaves notable gaps in clarity and completeness.
\end{itemize}

Overall, the reviewer seems to see this as valuable but incomplete work: interesting idea and useful dataset, but with methodological and experimental shortcomings that prevent it from being a clear accept. That aligns best with "6: marginally above the acceptance threshold.”

\vspace{0.3cm}

\textbf{Failure-mode flags:} superficial, inaccurate/hallucinatory, misguided

\textbf{Justification:}

\textbf{Superficial}

The review engages only shallowly with the paper’s main contributions. It focuses on high-level aspects (e.g., that the work studies multi-turn human jailbreaks and releases a dataset) but does not seriously discuss:

\begin{itemize}[nosep]
    \item Specific empirical findings (e.g., how multi-turn human attacks compare quantitatively to automated attacks across defenses).
    \item The proposed taxonomy and analysis of jailbreak tactics and failure modes.
    \item The detailed treatment of limitations, threat model assumptions, and evaluation caveats provided by the authors.
\end{itemize}

Instead, several weaknesses largely restate concerns that are already explicitly acknowledged and discussed in the paper (e.g., differences to automated attacks, variability in human red-team skill), without engaging with how the paper addresses or mitigates them. This makes the review feel surface-level relative to the richness of the manuscript.

\textbf{Inaccurate / hallucinatory}

The reviewer claims that there is "no discussion on plausible defenses for such multi-turn-human-based attack." This is inaccurate. The manuscript includes an explicit discussion of:

\begin{itemize}[nosep]
    \item Why current defenses, which are mostly tuned for single-turn settings, are insufficient for multi-turn adversarial interactions.
    \item How future defenses might need to adapt (e.g., post-training objectives that account for multi-turn interactions and longer contexts, and the implications of tool use and long-context agents).
\end{itemize}

While the paper may not introduce a new concrete defense algorithm, it \emph{does} discuss implications and directions for defense in this setting. Describing this as "no discussion" is therefore factually incorrect.

\textbf{Misguided}

Several criticisms are aimed at choices that are central, intentional aspects of the paper’s design rather than obvious shortcomings:

\begin{itemize}[nosep]
    \item The complaint that "only the Llama3-8b-Instruct model was used" treats the absence of multiple or stronger models (e.g., ChatGPT) as a major flaw, without acknowledging the practical cost of large-scale multi-turn human red-teaming and the paper’s primary goal: to compare \emph{defenses} under a fixed base model and realistic human-attack setting. The design choice is methodologically defensible given the problem the paper targets.
    \item The statement that the MHJ attack is "a soft-constrained version of attacks" and that the baselines differ in timing/budget is framed as a methodological weakness, but the paper explicitly defines a multi-turn human-chat threat model and explains why existing automated single-turn attacks are not directly comparable. Criticizing this as though the mismatch were an oversight rather than a deliberate focus is somewhat misplaced.
    \item The concern about variability in red-teamer skill is presented as a reason to prefer automated attacks, but the manuscript itself highlights this limitation, motivates the use of multiple experienced red teamers, and releases the MHJ dataset precisely to help the community move toward more standardized and automatable evaluations. Treating this as an unaddressed problem rather than an explicitly discussed trade-off underestimates the authors’ own framing.
\end{itemize}

Overall, the review is not especially harsh or lenient, but it does contain one clear factual inaccuracy and critiques some well-justified design decisions as if they were simple omissions, while remaining relatively superficial about key contributions of the paper.

\vspace{0.3cm}
\hrule
\vspace{0.3cm}

{\large \textbf{Reviewer \#6}}

\textbf{Summary:}

This paper focuses on the safety vulnerabilities of LLMs in multi-turn conversations. The authors designed a comprehensive human red teaming pipeline, including "Attempt Jailbreak" and "Validate Jailbreak," and found that LLMs exhibit more safety risks in multi-turn dialogue conversations compared to single-turn automated attacks.

\textbf{Strengths:}
\begin{enumerate}[nosep]
    \item Most previous papers have focused on LLM jailbreaking in single-turn conversations, while this work emphasizes multi-turn conversations, presenting a new scenario.
    \item The red-teaming results show that multi-turn human jailbreaks outperform current automated attacks, which exposes more safety vulnerabilities in LLMs.
    \item The authors have also released their Multi-turn Human Jailbreaks dataset, which can support further research on multi-turn jailbreaks.
\end{enumerate}

\textbf{Weaknesses:}
\begin{enumerate}[nosep]
    \item The human red-teaming was conducted only on Llama; further evaluations on other LLMs would more comprehensively illustrate the safety vulnerabilities in multi-turn conversations. Additionally, the human red-teaming data collected on Llama could potentially be used to examine safety issues in other LLMs under multi-turn scenarios.
    \item A more detailed analysis of the effectiveness of different tactics could provide deeper insights into multi-turn jailbreaks.
\end{enumerate}

\textbf{Questions:}
NA

\vspace{0.3cm}

\textbf{Rating:} 8: accept, good paper

\textbf{Justification:}

The review describes a solid and meaningful \textbf{contribution} with clear novelty and practical value:
\begin{itemize}[nosep]
    \item Novelty and relevance: The paper moves beyond the well-studied single-turn jailbreak setting and systematically focuses on multi-turn conversations, which is an important and underexplored but realistic scenario for LLM safety.
    \item Methodological contribution: The authors design a comprehensive human red-teaming pipeline (with explicit Attempt Jailbreak" and Validate Jailbreak" stages), suggesting a thoughtful and structured experimental setup rather than ad-hoc red-teaming.
    \item Empirical significance: The finding that multi-turn human jailbreaks outperform current automated attacks indicates that existing defenses are insufficient in realistic interaction patterns and highlights practically important safety vulnerabilities.
    \item Resource contribution: The release of the Multi-turn Human Jailbreaks dataset is a concrete asset that can enable and accelerate further research on multi-turn jailbreaks.
\end{itemize}

The \textbf{weaknesses} raised are important but appear incremental rather than fatal:
\begin{itemize}[nosep]
    \item Limited model coverage: Experiments focus only on Llama; broader evaluation across multiple LLMs would strengthen the generality of the conclusions, but the review does not indicate that this limitation invalidates the core findings.
    \item Depth of analysis: The reviewer asks for more detailed analysis of different tactics. This suggests room for improvement in insight and interpretability, but not a fundamental flaw in the methodology or results.
\end{itemize}
Overall, the review portrays the work as a good paper with clear novelty, solid methodology, and useful artifacts, with limitations that are reasonable for a conference submission rather than reasons for rejection. Hence, a score of 8 (accept, good paper) is appropriate.

\vspace{0.3cm}

\textbf{Failure-mode flags:} superficial.

\textbf{Justification:}

\begin{enumerate}[nosep]
    \item The review remains at a high level and does not engage deeply with methodological details. For example, it mentions a "comprehensive human red teaming pipeline” with stages like "Attempt Jailbreak” and "Validate Jailbreak,” but provides no discussion of how these stages are designed, what criteria are used, or any potential limitations or biases of this setup.
    \item The strengths section largely restates the paper’s apparent contributions (focus on multi-turn conversations, better performance of human jailbreaks, release of a dataset) without critically assessing their significance, novelty, or empirical robustness.
    \item The weaknesses identified are quite generic. Saying that red-teaming only on Llama is limiting and that "more detailed analysis of the effectiveness of different tactics” would be useful are reasonable points, but the review does not specify what concrete additional experiments or analyses would meaningfully improve the work, nor does it probe any deeper conceptual or empirical shortcomings.
    \item Overall, the review reads as a brief summary plus a couple of high-level comments, rather than a thorough evaluation of experimental design, analysis, or positioning relative to prior work, which is characteristic of a somewhat superficial review.
\end{enumerate}

\vspace{0.3cm}
\hrule
\vspace{0.3cm}

{\large \textbf{Reviewer \#7}}

\textbf{Summary:}

This paper exposes the significant safety vulnerabilities of Large Language Models (LLMs) in multi-turn dialogue by conducting human red-teaming. By exploiting a diverse of human tactics, human red-teamers achieve a 70\% attack success rate on Harmbench against strong defense baselines. This paper finally compiles these results into the Multi-Turn Human Jailbreaks (MHJ) dataset.

\textbf{Strengths:}
\begin{itemize}[nosep]
    \item \textbf{Valuable Findings, Including an Open-Source Jailbreak Dataset:} This paper provides a valuable contribution by conducting the first multi-turn human red-teaming experiments, revealing significant safety vulnerabilities that bypass current state-of-the-art defense methods. Given that existing safety alignment techniques primarily target single-turn jailbreaks, these findings effectively motivate the community to develop more robust defenses capable of countering multi-turn attacks.
\end{itemize}

\textbf{Weaknesses:}
\begin{itemize}[nosep]
    \item \textbf{Lack of Strong Automated Attack Baselines:} Some black-box attacks, such as CodeAttack [1] and PAP [2], are not included in the experiments, despite utilizing similar "Request Framing” tactics as those in this paper. Including these baselines in future comparisons would offer a clearer perspective on the relative effectiveness of this paper's approach.
    \item \textbf{Unclear Evaluation Metric for Model Unlearning:} This paper proposes manual review to assess attack outcomes in model unlearning experiments, yet does not clearly define what constitutes a successful attack. For example, does success mean the model output includes content from the forget set? A more detailed description of the evaluation metric for model unlearning experiments would improve clarity.
\end{itemize}
\begin{enumerate}[nosep]
    \item \emph{CodeAttack: Revealing Safety Generalization Challenges of Large Language Models via Code Completion.}
    \item \emph{How Johnny Can Persuade LLMs to Jailbreak: Rethinking Persuasion to Challenge AI Safety by Humanizing LLMs.}
\end{enumerate}

\textbf{Questions:}
\begin{itemize}[nosep]
    \item \textbf{Counter-Intuitive Experimental Results:} The effectiveness of the "Direct Request” tactic used by human red-teamers, as shown in Figure 8, is unexpected. In this paper, "Direct Request” refers to directly asking the LLM to produce harmful content, which should logically not be highly effective against safety defenses. An analysis of how various tactics influence the attack success rate would be beneficial, as it could offer insights into the specific failure modes of these models.
    \item \textbf{Lack of In-Depth Analysis on the High Effectiveness of Human Multi-Turn Jailbreaks:} Many of the tactics employed by human red-teamers, such as "Injection,” "Obfuscation,” "Output Format,” and "Request Framing,” are also used by automated attacks. Are there particular factors that enable human-led jailbreaks to outperform automated attacks using similar tactics? Identifying these factors could provide valuable insights into the unique strengths of human-driven attacks.
\end{itemize}
I would consider raising my score if the authors address these concerns and questions.

\vspace{0.3cm}

\textbf{Rating:} 6: marginally above the acceptance threshold

\textbf{Justification:}

The review is overall positive about the core contribution: it highlights that the paper is (i) the first to conduct multi-turn human red-teaming experiments on LLMs, (ii) exposes substantial safety vulnerabilities with a high attack success rate (70\% on HarmBench), and (iii) releases a useful open-source dataset (MHJ). These points indicate that the reviewer sees the work as novel, impactful, and relevant to the community.

However, the reviewer also identifies non-trivial weaknesses. They point out missing strong automated black-box baselines (CodeAttack, PAP), which limits the strength of the empirical comparison, and they raise concerns about the clarity of the evaluation metric for model unlearning. In addition, the questions section emphasizes that some of the experimental findings (e.g., the high effectiveness of direct requests) are under-analyzed, and that the mechanisms behind the superiority of human multi-turn attacks over automated ones are not sufficiently explored. The reviewer explicitly states that they would consider \emph{raising} their score if these concerns were addressed, suggesting that their current stance is cautiously positive rather than fully enthusiastic.

Taken together, this combination of a clearly valuable main contribution with methodological and analytical gaps is most consistent with a score of 6 (marginally above the acceptance threshold): the paper is viewed as worth accepting, but with reservations that prevent it from reaching the level of a clear "accept, good paper” (8) or higher.

\vspace{0.3cm}

\textbf{Failure-mode flags:} inaccurate/hallucinatory, superficial

\textbf{Justification:}

\textbf{Inaccurate / hallucinatory}
\begin{itemize}[nosep]
    \item The review claims that the evaluation metric for the unlearning experiments is "not clearly defined" and asks, for example, whether success means "the model output includes content from the forget set." However, the paper does spell this out: for WMDP-Bio, they use the ground truth answer corresponding to each prompt to verify the attack, and define success as the model output semantically containing all the information in the label, as judged by a human reviewer. Thus, the alleged lack of clarity is factually incorrect.
    \item The strength section claims this paper "provides \ldots the first multi-turn human red-teaming experiments." The paper instead positions its work as examining a realistic but underexplored multi-turn threat model and notes that large-scale human red teaming is rare outside industry, but it does not assert a clean "first" in the way the reviewer phrased it. This is a mild overstatement of novelty not actually grounded in the manuscript.
\end{itemize}

\textbf{Superficial}
\begin{itemize}[nosep]
    \item The criticism of an "unclear evaluation metric" for model unlearning ignores the dedicated description of how success is judged in the WMDP-Bio setting (a semantic match to ground-truth answers via human review). This suggests the reviewer did not fully engage with the relevant part of the methodology.
    \item The "lack of strong automated attack baselines" complaint is framed entirely around the absence of two specific attacks (CodeAttack and PAP), without engaging with the existing suite of six automated attacks (including white-box gradient-based methods) and the ensemble upper-bound attack success rate (ASR) the authors already use, nor with their explicit threat-model-driven exclusions (e.g., Crescendo due to missing code, or attacks that modify model internals). This makes the baseline critique feel somewhat shallow relative to the paper’s fairly detailed experimental setup.
    \item The review does not really grapple with the paper’s central design choices and limitations (multi-turn human pipeline, validation stages, conservatively biased ASR estimates, and discussion of differences between human and automated pipelines) that occupy multiple sections and appendices. Instead, it focuses on a small number of surface-level points (missing two baselines, asking for more analysis of one tactic) and offers only a single, high-level strength. This imbalance contributes to the overall superficiality.
\end{itemize}

\vspace{0.3cm}
\hrule
\vspace{0.3cm}

{\large \textbf{Reviewer \#8}}

\textbf{Summary:}

This paper looks at red teaming via multi-turn human interaction and releases an accompanying dataset of the jailbreaks. This is a useful study: the vast majority of the research effort and available data is focused on single-turn attacks, with only a few exceptions (e.g. Anthropic's hh-rlhf dataset having harmful multi turn conversations). Overall, the paper is clear and well written, and supplies useful well curated data for future work.

\textbf{Strengths:}
\begin{itemize}[nosep]
    \item Human driven red teaming data, particularly multi-turn due to its scarcity, is always useful. There are many synthetically generated datasets, but high quality human driven attacks are a valuable resource. In particular, each prompt here had a high degree of manual curation (in comparison to datasets like hackaprompt which contain many low quality samples).
    \item The analysis across a few defenses is useful, and does highlight that even SOTA defenses on highly studied benchmark harmful questions can be broken in around 15min on a large subset of questions (e.g. often quicker than algorithmic attack runtime).
\end{itemize}

\textbf{Weaknesses:}
\begin{itemize}[nosep]
    \item It would have been useful to see results which have human single turn jailbreak attempts in Figure 3. At the moment, the "Human" attacks have two variables changed compared to the others: the attack source (e.g. handcrafted), and additionally have multi-turn capabilities. Hence it makes it challenging to disambiguate if the difference in performance is due to the multi-turn aspect, or if humans given enough time remain better than automated based methods at creating jailbreaks.
    \item I am unclear as to why results against Cygent defense could not be carried out in the same setup as the original paper: the Llama model is open source, and the defense has a published paper. It would have enabled stronger reproducibility and clearer interpretation of results.
    \item Releasing the non-successful jailbreak attempts as well can be beneficial as it is still a useful resource, for example as training/fine-tuning data or to carry out further analysis.
    \item Although different styles of harmfulness were investigated: both "regular" harmbench style questions, but also WMDP-Bio for different attack objectives and domain performance from looking into the supplementary material it seems like just the harmbench data was released. Given the dataset is the core contribution of the paper it would have been useful to include the other domain data.
\end{itemize}

\textbf{Questions:}

Are there statistics on how many red-team members there are/distributions on number of samples provided per red-team member?

\vspace{0.3cm}

\textbf{Rating:} 8: accept, good paper

\textbf{Justification:}

\begin{itemize}[nosep]
    \item The review describes the work as a \emph{useful study}, highlighting that most prior work and data focus on single-turn attacks, whereas this paper targets the underexplored setting of \emph{multi-turn human red teaming} and releases a corresponding dataset. This indicates a clear and meaningful contribution rather than an incremental tweak.
    \item The dataset is portrayed as \emph{high-quality} and \emph{well curated}, with an explicit comparison to synthetically generated or lower-quality resources (e.g., hackaprompt). This suggests strong lasting value to the community.
    \item The analysis of multiple defenses and the empirical finding that even SOTA defenses can be broken in roughly 15 minutes are framed as \emph{useful} and informative, reinforcing the paper’s impact on understanding current safety limitations.
    \item The weaknesses raised (lack of a human single-turn baseline for clearer attribution to multi-turn effects, missing Cygent reproduction details, omission of unsuccessful attempts, and incomplete release of all domain data) are presented as limitations or missed opportunities for strengthening the work, but not as fundamental flaws that would undermine the main results.
    \item Overall, the tone is clearly positive and confident in the paper’s value and clarity, but not enthusiastic enough (nor claiming sufficiently broad or groundbreaking impact) to justify a "strong accept." This aligns best with an \emph{accept, good paper} rather than merely marginal acceptance.
\end{itemize}

\vspace{0.3cm}

\textbf{Failure-mode flags:} Inaccurate / hallucinatory (minor).

\textbf{Justification:}

The review’s second weakness claims it is unclear why experiments on Cygent/CYGNET" could not be reproduced in the same setup as the original paper, arguing that the Llama model is open source, and the defense has a published paper." However, the manuscript explicitly states that CYGNET is \emph{closed-source/proprietary} and that the authors conducted human red teaming via an external API (Gray Swan AI), while also noting they did not run automated attacks on CYGNET in their own setup and instead cite automated results from the original work (with caveats about comparability). This makes the review’s implication that straightforward reproduction should have been possible materially inconsistent with the manuscript’s described experimental constraints.

\end{fullsummarybox}
%--------------------------------------------------

\begin{metareviewbox}
\textbf{Recommendation:} Reject

\textbf{Meta Review:}

Strength:
\begin{enumerate}[nosep]
    \item Most previous papers have focused on LLM jailbreaking in single-turn conversations, while this work emphasizes multi-turn conversations, presenting a new scenario of evaluating the LLM jailbreak robustness from a different perspective.
    \item The authors have also released their Multi-turn Human Jailbreaks dataset, which can support further research on multi-turn jailbreaks.
\end{enumerate} 
Weakness:
\begin{enumerate}[nosep]
    \item This paper appears to be strictly an experimental investigation (based on unfair comparison) rather than a proposal of novel techniques.
    \item Limited evaluations: limited to llama-3-8b-instruct, limited set of evaluated defenses, and unclear evaluation metric for model unlearning.
    \item Lack of more detailed analysis of the effectiveness of different tactics in providing deeper insights into multi-turn jailbreaks. 
\end{enumerate} 
While this work contributes to the study of Multi-turn Human Jailbreaks, it cannot be accepted at its current status due to the main weaknesses, including limited evaluations and lack of methodological analyses.
\end{metareviewbox}

%%%%%%%%%%%%%%%%%%%%%%%%%%%%%%%%%%%%%%%%%%%%%%%%%%%%%%%%%%%%

% \newpage
% \input{checklist.tex}

\end{document}